\documentclass[review,sort&compress]{elsarticle}

\usepackage{amssymb}
\usepackage{epsfig}
\usepackage{graphics}
\usepackage{graphicx}
\usepackage{subfigure}
\usepackage{exscale}
\usepackage{latexsym}
\usepackage{makeidx}
\usepackage[section]{placeins}
\usepackage{xtab}
\usepackage{epsf}
\usepackage{bbm}

\usepackage{lineno}

\usepackage{xcolor}

\journal{Nuclear Instruments and Methods in Physics Research Section A}

\begin{document}
\begin{frontmatter}

\title{Measurement of photo- and radio-luminescence of thin ThF$_4$ films}

\author[INFN-GE]{Mikhail Osipenko\corref{cor1}}
\ead{osipenko@ge.infn.it}
\author[JRC-KH]{Carlos de Almeida Carrapico}
\author[JRC-KH]{Dmytro Burdeinyi}
\author[INFN-GE]{Roberto Caciuffo}
\author[JRC-KH]{Rachel Eloirdi}
\author[INFN-GE,UNI-GE]{Mauro Giovannini}
\author[JRC-KH]{Alban Kellerbauer}
\author[JRC-KH]{Rikard Malmbeck}
\author[INFN-GE]{Marco Ripani}
\author[INFN-GE,UNI-GE]{Mauro Taiuti}

\cortext[cor1]{Please address correspondence to Mikhail Osipenko}

\address[JRC-KH]{JRC, Karlsruhe, 76125 Germany}
\address[INFN-GE]{INFN, Sezione di Genova, Genova, 16146 Italy}
\address[UNI-GE]{Universit\`a  di Genova, Genova, 16146 Italy}

\begin{abstract}
We conducted measurements on the photo- and radio-luminescence of thin ThF$_4$ films in both the UV and visible ranges. In the UV range, we found that both luminescences are at a similar level as the internal dark counting noise of the photo-multiplier-tube (PMT). Our results suggest that thin ThF$_4$ crystals could be used as a target for the search for $^{229m}$Th and as a medium for the future nuclear clock. The measurements indicate that using a small and thin ThF$_4$ film can reduce background noise while maintaining the signal at the same level, achieved by increasing the $^{229}$Th enrichment. Our developed apparatus is now ready for direct measurements of $^{229m}$Th excitation and decay in ThF$_4$.

\end{abstract}


\end{frontmatter}

\section{Introduction}\label{sec:intro}
$^{229m}$Th is the lowest known nuclear isomeric state. Its energy of 8.3~eV~\cite{Seiferle} is accessible with the currently available highly monochromatic lasers, opening the way to several practical applications. In particular, it enables the realization of a so-called nuclear clock~\cite{Tkalya_photoexcitation,peik}. However, in order for the precision of a nuclear clock to surpass that of existing atomic optical clocks, Internal Conversion (IC), which is the dominant $^{229m}$Th decay channel~\cite{Wense,Seiferle_lifetime}, must be suppressed. The suppression of IC decay was predicted in Refs.~\cite{Tkalya_n3_dep,Tkalya_n3_dep_prc} and recently was observed in ionized $^{229}$Th atoms~\cite{Yamaguchi_tau_th229m_vacuum} as well as in $^{229}$Th dispersed in solid matrices with a large bandgap~\cite{Tiedau_tau_th229m_caf2,Kraemer_tau_th229m_mgf2}. Crystals used to search for $^{229m}$Th in a solid matrix include CaF$_2$\cite{Tiedau_tau_th229m_caf2}, MgF$_2$\cite{Kraemer_tau_th229m_mgf2}, and LiSrAlF$_6$\cite{Elwell_tau_th229m_licaf}. The crystals were all grown to sizes greater than 1~mm$^3$, which increased the background radio-luminescence due to their large active volume that could fully absorb the $\alpha$-particles of the $^{229}$Th decay~\cite{Tiedau_tau_th229m_caf2}. While the successful observation of $^{229m}$Th has been accomplished in these crystals, the efficiency of the detection process has not been determined. Furthermore, the refractive indices $n$ of the used crystals are very similar, which prevents precise testing of the theoretical $n^3$-dependence of the signal \cite{Tkalya}. We speculate that for the thin films with thickness smaller than the wavelength of emitted $\gamma$ the $^{229m}$Th lifetime could depend on the film thickness.

In our current study, we are analyzing the photo-luminescence and radio-luminescence of ThF$_4$ thin films. ThF$_4$ is a material that is considered an excellent candidate as a working matrix for a $^{229m}$Th nuclear clock. This material naturally contains thorium, which helps avoid band distortions and lattice defects when doped with $^{229}$Th.  Additionally, ThF$_4$ is an insulator with a bandgap of 10.2(2) eV~\cite{band_gap_thf4}, wide enough to suppress $^{229m}$Th IC decay.  ThF$_4$ can be easily grown in the form of high-quality thin films and its refractive index $n$ at 150~nm is approximately 1.826~\cite{thf4_refractive_index}, which is significantly different from those of the other crystals tested so far (1.485 for LiSrAlF$_6$, 1.488 for MgF$_2$ and 1.586 for CaF$_2$). Before searching for $^{229m}$Th excitation and $\gamma$-decay in ThF$_4$, it is essential to characterize the major background sources, namely photo- and radio-luminescence. 

In section~\ref{sec:setup}, we will describe the experimental apparatus that we have designed and built, as well as the characteristics of the ThF$_4$ thin layers. In section~\ref{sec:photo_mc_sim}, we will explain the specific configuration for the photo-luminescence measurements and its simulations. The photo-luminescence data will be presented in section~\ref{sec:photo_lum_exp}. A similar description will be provided for radio-luminescence measurements in sections~\ref{sec:radio_mc_sim} and~\ref{sec:radio_lum_exp}. The interpretation of the obtained quantities compared to the expected $^{229m}$Th decay signal will be given in section~\ref{sec:projection_expect}, followed by the conclusions in section~\ref{sec:conclusions}.

\section{Experimental setup}\label{sec:setup}

\subsection{VUV-photon source}\label{sec:source}
\begin{figure}[!ht]
\begin{center}
\includegraphics[bb=1cm 2.5cm 20cm 16cm, scale=0.25, angle=270]{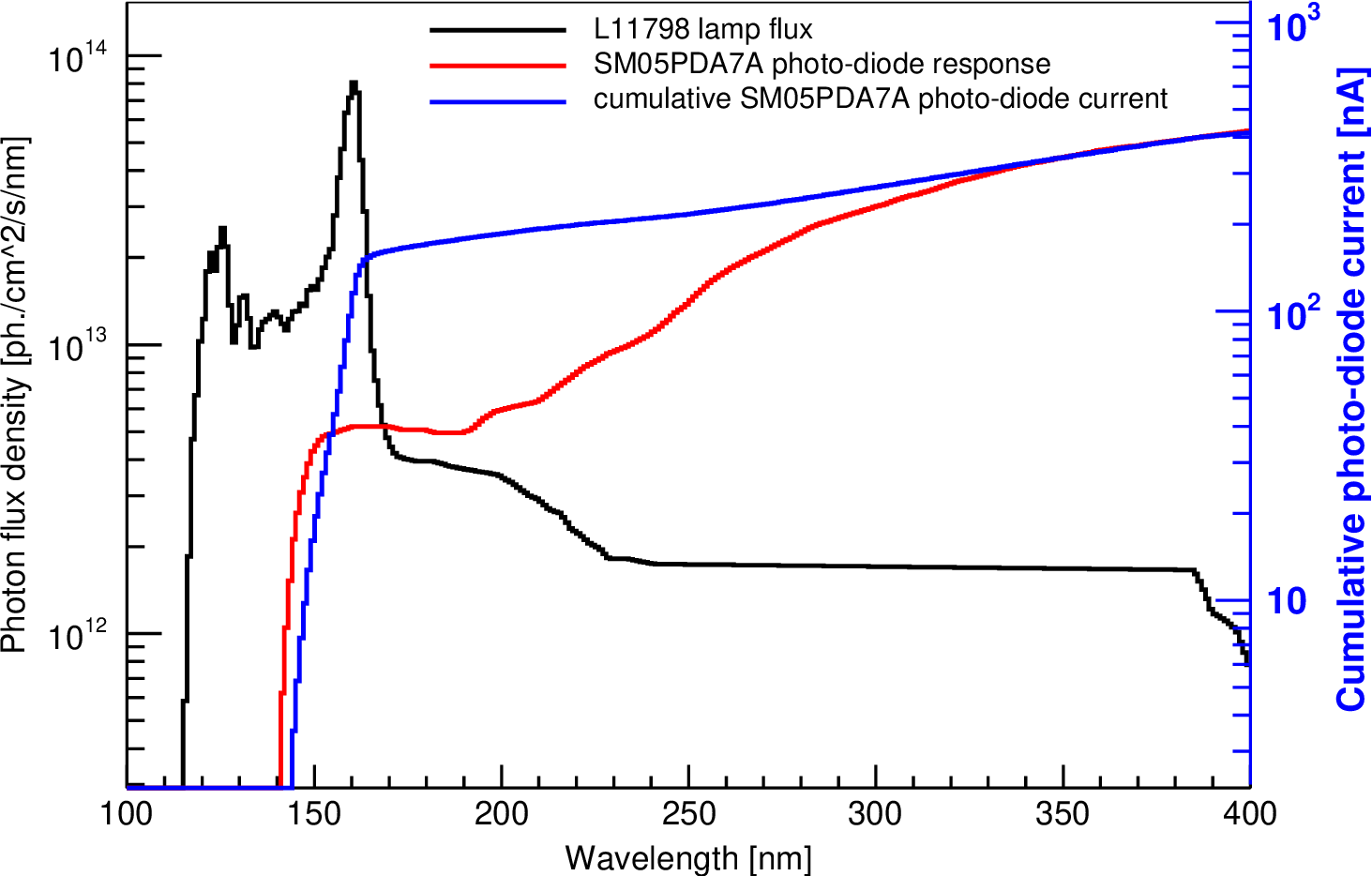}
\caption{\label{fig:source_flux} Photon flux from the L11798 lamp measured by the SM05PDA7A photo-diode at the sample location. The emission spectrum was taken from Hamamatsu datasheets and normalized to the observed diode current. The SM05PDA7A photo-diode response and its cumulative current are shown for comparison.}
\end{center}
\end{figure}

The experimental setup for photo-luminescence measurements is based on the vacuum ultraviolet (VUV) light source H$_2$D$_2$-lamp L11798 of Hamamatsu~\cite{L11798_datasheets}. The lamp has a broad spectrum, peaked in the VUV range, shown in Fig.~\ref{fig:source_flux}. 

\begin{figure}
\begin{center}
\includegraphics[scale=0.45]{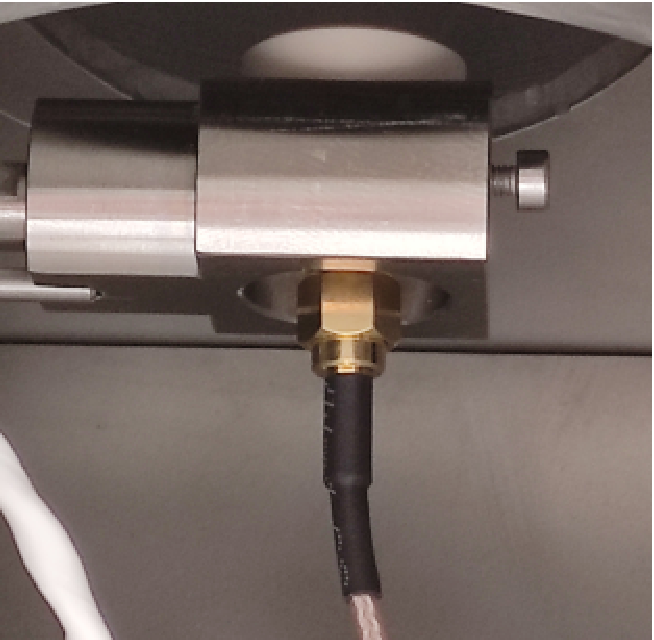}
\caption{\label{fig:diode_foto} Picture of the photo-diode holder installed under the VUV lamp. The coaxial cable connects the photo-diode to its readout controller.}
\end{center}
\end{figure}

\begin{figure}[!ht]
\begin{center}
\includegraphics[bb=1cm 2.5cm 20cm 22cm, scale=0.25, angle=270]{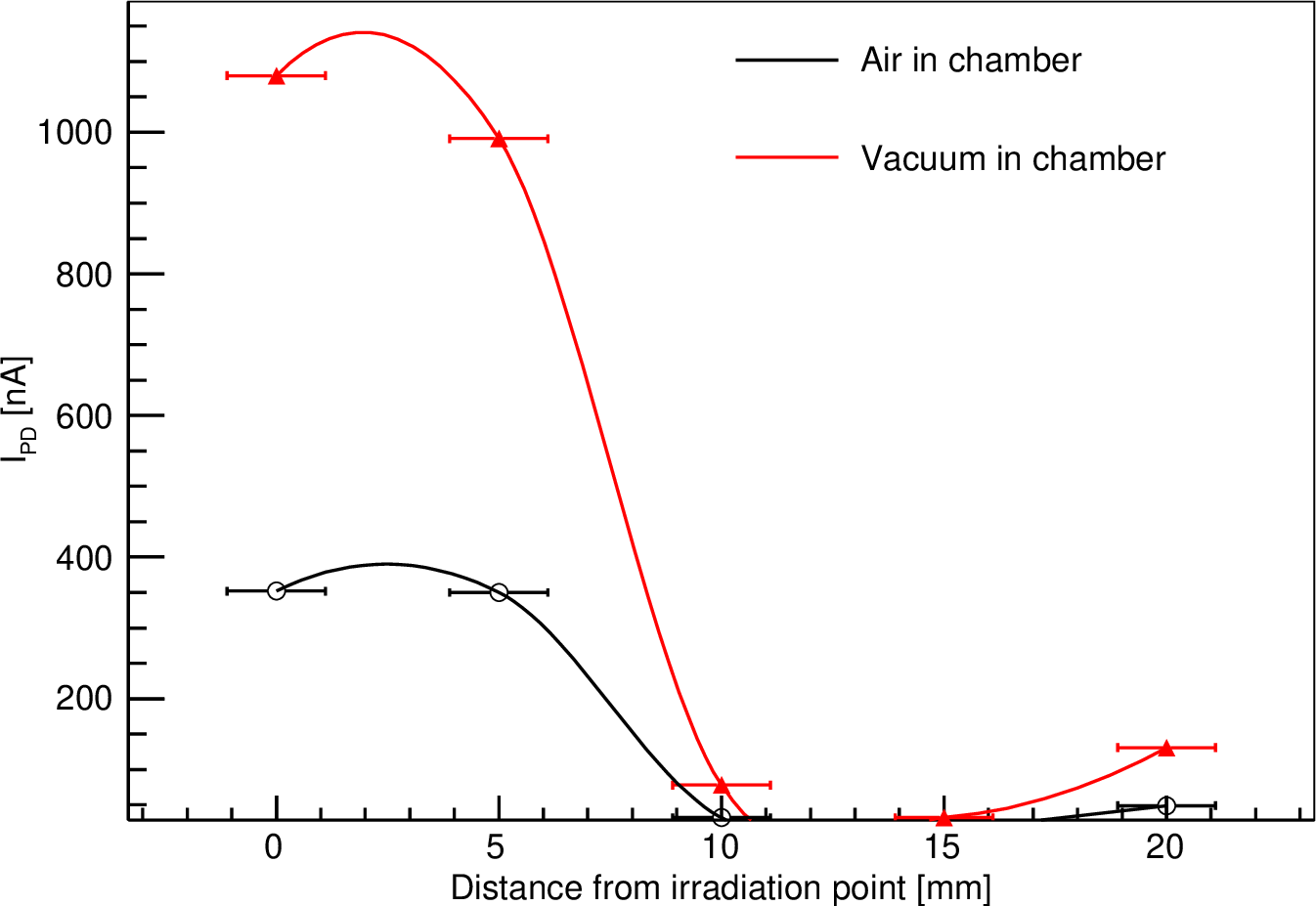}
\caption{\label{fig:photo_diode_current} Electrical current measured by the SM05PDA7A photo-diode at multiple points along the translation axis, away from the central irradiation point. The measurements were performed under vacuum and in air to estimate the VUV component of the measured signal.}
\end{center}
\end{figure}

To normalize the known lamp emission spectrum~\cite{L11798_datasheets}, we performed a series of measurements using a VUV-sensitive photo-diode SM05PDA7A~\cite{SM05PDA7A_datasheets} installed in the location of the irradiated sample. Fig.~\ref{fig:diode_foto} shows the picture of the experimental setup. Moving this diode along the translation stick axis from the central irradiation point, we measured the photon beam profile shown in Fig.~\ref{fig:photo_diode_current}. The photon beam spot had a full width at half maximum FWHM = 15 mm whilst the peak current of the photo-diode was 1080~nA under vacuum and 352~nA in air. Thus, in air, we observed only 33\% of the light signal, due to absorption of the VUV part of the lamp emission spectrum.

The comparison of the lamp's emission spectrum with the photo-diode sensitivity, as shown in Fig.~\ref{fig:source_flux}, demonstrates that the primary contribution to the observed current comes from the peak around 161~nm (with a FWHM of 5~nm), where the photo-diode sensitivity is approximately 0.01~A/W. To estimate the total photon flux at the measured point, a simplified calculation was used with the following equation:

\begin{equation}\label{eq:photon_flux}
I_{pd} [nA] = \phi_\gamma \Biggl[\frac{ph}{cm^2~s~nm}\Biggr] \times R_{pd} \Biggl[\frac{A}{W}=\frac{nA~s}{nJ}\Biggr] \times S_{pd} \Bigl[cm^2\Bigr] \times E_\gamma \Biggl[\frac{nJ}{ph}\Biggr] \times \Delta\lambda \Bigl[nm\Bigr] ~,
\end{equation}
\noindent where $I_{pd}$ is the measured photo-diode current, $\phi_\gamma$ is the differential photon flux, $R_{pd}$ is the wavelength dependent photo-diode response, $S_{pd}$ is the photo-diode effective area, and $E_\gamma$ is the photon energy. The photo-diode active region had a 2.2~mm side, thus $S_{pd}=0.0484$~cm$^2$. In the region around 161~nm $R_{pd}=0.01$~nA/nW, $E_\gamma=7.7$~eV=12.3$\times 10^{-10}$ nJ/ph and $\Delta\lambda=5$~nm. Combining all these numbers we estimate the photon flux in the main peak $\phi_\gamma = 1.8\times 10^{14} ph/(cm^{2}~s~nm$).
%
%
\noindent Fig.~\ref{fig:source_flux} shows the results of a more careful assessment based on a bin-by-bin convolution. In this analysis, we have normalized the datasheet emission spectrum of the lamp to the observed photo-diode current. The fraction of the photo-diode current above 200~nm (air absorption) was estimated to be 55\%, which is significantly higher than the observed 33\%.

\subsection{Readout electronics and DAQ}\label{cap:elec}
The PMT anode produces very fast direct signals. For instance, for R6835 we have measured signals with 3~ns rise and fall times~\cite{R6835_datasheets}. This implies that a fast readout requires about 500~MHz analogue bandwidth. However, the expected signal is due to single photo-electrons, which generate roughly 0.1~pC (R6835 at 2500~V) charge on the PMT anode, translating to 2~mV peaks in a 50~Ohm load. The fast electronics generate competing noise in the order of a few mV/GHz, and there is significant electromagnetic interference (EMI) in this frequency domain. This makes it challenging to trigger signals below the single photo-electron amplitude and to differentiate them from signals generated by a few photo-electrons.  Additionally, our measurements do not have fast coincidences, which could help improve noise rejection in a fast readout scheme. Consequently, we have opted to use the slow (1~MHz) charge-sensitive readout scheme. Reducing the bandwidth by three orders of magnitude enables much higher resolution and better noise rejection. In this scheme, EMI noise is not expected to be significant, thereby minimizing unwanted and uncontrollable electronic noise. This scheme is based on the charge-sensitive preamplifier Ortec-113~\cite{ortec113}, connected to the PMT anode and the DT5730 digitizer featuring the DPP-PHA firmware~\cite{DPP-PHA}.

The Ortec-113 is a parasitic capacitance preamplifier with about 45~pF internal capacitance. The rise and fall times of its signal are less than 20 ns and 50~$\mu$s, respectively. With an integral non-linearity smaller than $\pm$0.02\%, a temperature coefficient of $\pm$0.01\%/$^\circ$C in the range 0 to 50 $^\circ$C, and a noise of $<$0.1 mV RMS at the output, the Ortec-113 enables very stable measurements of low-rate, small-amplitude events. The signals from the preamplifier were acquired by the CAEN DT5730 digitizer~\cite{DT5730}, recording pulse heights for each event reconstructed by the trapezoidal filter of the DPP-PHA firmware~\cite{DPP-PHA}. For these measurements, we utilized the standard CAEN COMPASS Data Acquisition System (DAQ)~\cite{CAEN-COMPASS}. To optimize the signal-to-noise ratio and charge resolution, we used the following configurations of COMPASS/DT5730: 

\begin{itemize}
\item coarse gain $\times$4,
\item risetime 16~ns, corresponding to the observed Ortec~113 output signals,
\item trigger threshold 5 lsb,
\item fast discriminator smoothing 128 samples,
\item trigger holdoff 8.2~$\mu$s,
\item trapezoid risetime of 5~$\mu$s,
\item trapezoid top of 1~$\mu$s,
\item averaged peak samples 64,
\item peak holdoff 8~$\mu$s.
\end{itemize}

The digitized signal PH in ADC channels can be used to estimate the absolute charge collected on the PMT anode by multiplying it by (0.5~V/16384 ch)$\times$4$\times$45~pF=5.5$\times$10$^{-3}$~pC/ch, where the factor 4 comes from the Ortec~113 output and the digitizer input impedance matchings to 50~Ohm.

\subsection{ThF$_4$ samples}\label{cap:thf4_samples}
\begin{figure}[h]
\begin{center}
\includegraphics[bb=0cm 0cm 16cm 16cm, scale=0.35]{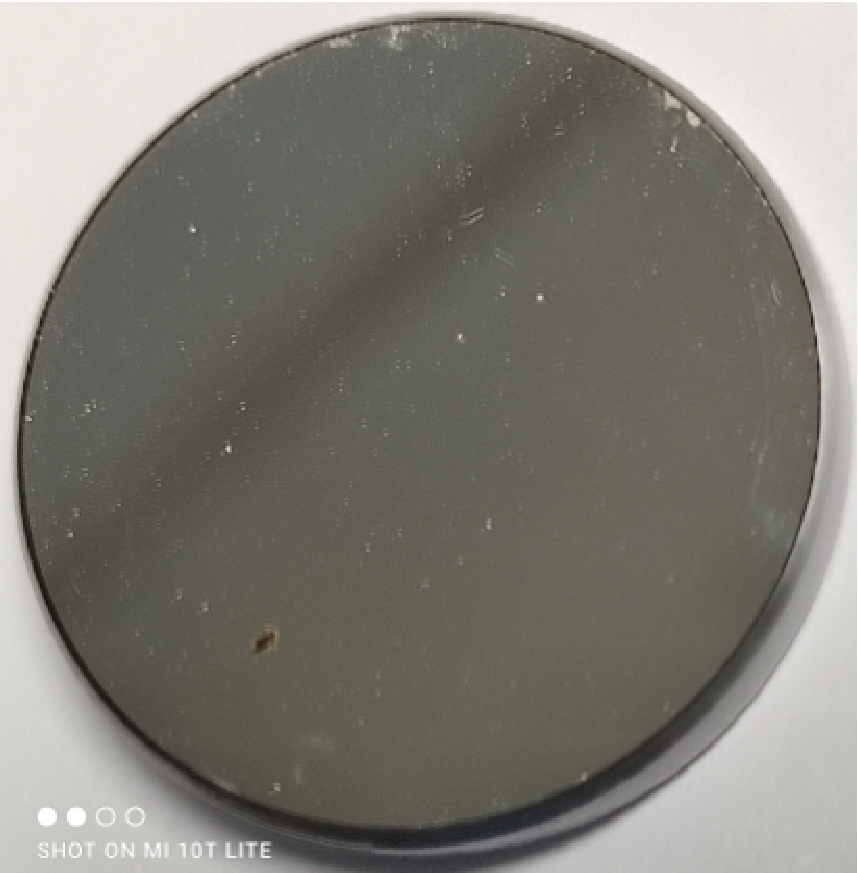}~%
\includegraphics[bb=0cm 0.4cm 4cm 5cm, scale=1.6]{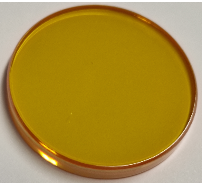}
\caption{\label{fig:si_znse_thf4} II-VI GmbH silicon mirror (left) and ZnSe partial reflector (right) with a ThF$_4$ external layer on one side.}
\end{center}
\end{figure}

We have tested two commercial samples of optical elements featuring an amorphous ThF$_4$ film at their outermost layer. The elements, shown in Fig.~\ref{fig:si_znse_thf4}, are produced by II-VI GmbH~\cite{II-VI}.



%

The first sample is a 3.05 mm thick silicon mirror of 1-inch diameter and a plane surface. One side of the mirror is opaque, while the 
other side is a so-called total reflector (see Fig.~\ref{fig:si_znse_thf4}, left) with an outermost 300-nm-thick layer made of amorphous ThF$_4$, which has high reflectivity at 10.6~$\mu$m.There are additional coating layers between the external ThF$_4$ layer and the silicon substrate, but for our purposes, we assume that their contribution to the radio- and photo-luminescence of the sample is negligible.

%
%
%
%
%
%

The second sample is a 3.05 mm thick ZnSe lens of 1-inch diameter. The ZnSe substrate is semi-transparent and has a yellow colour. One side of the lens acts as a partial reflector with an outermost polarization converter (PS) layer featuring high reflectivity at 10.6~$\mu$m. This PS layer is made of amorphous ThF$_4$ and has a thickness of 200~nm. Beneath the external ThF$_4$ layer and above the ZnSe substrate, there are additional coating layers, although they are not specified. Upon visual inspection, there is no discernible difference between the two sides of this sample.

\section{Photo-luminescence setup and Monte Carlo simulations}\label{sec:photo_mc_sim}
\begin{figure}
\begin{center}
\includegraphics[scale=0.50]{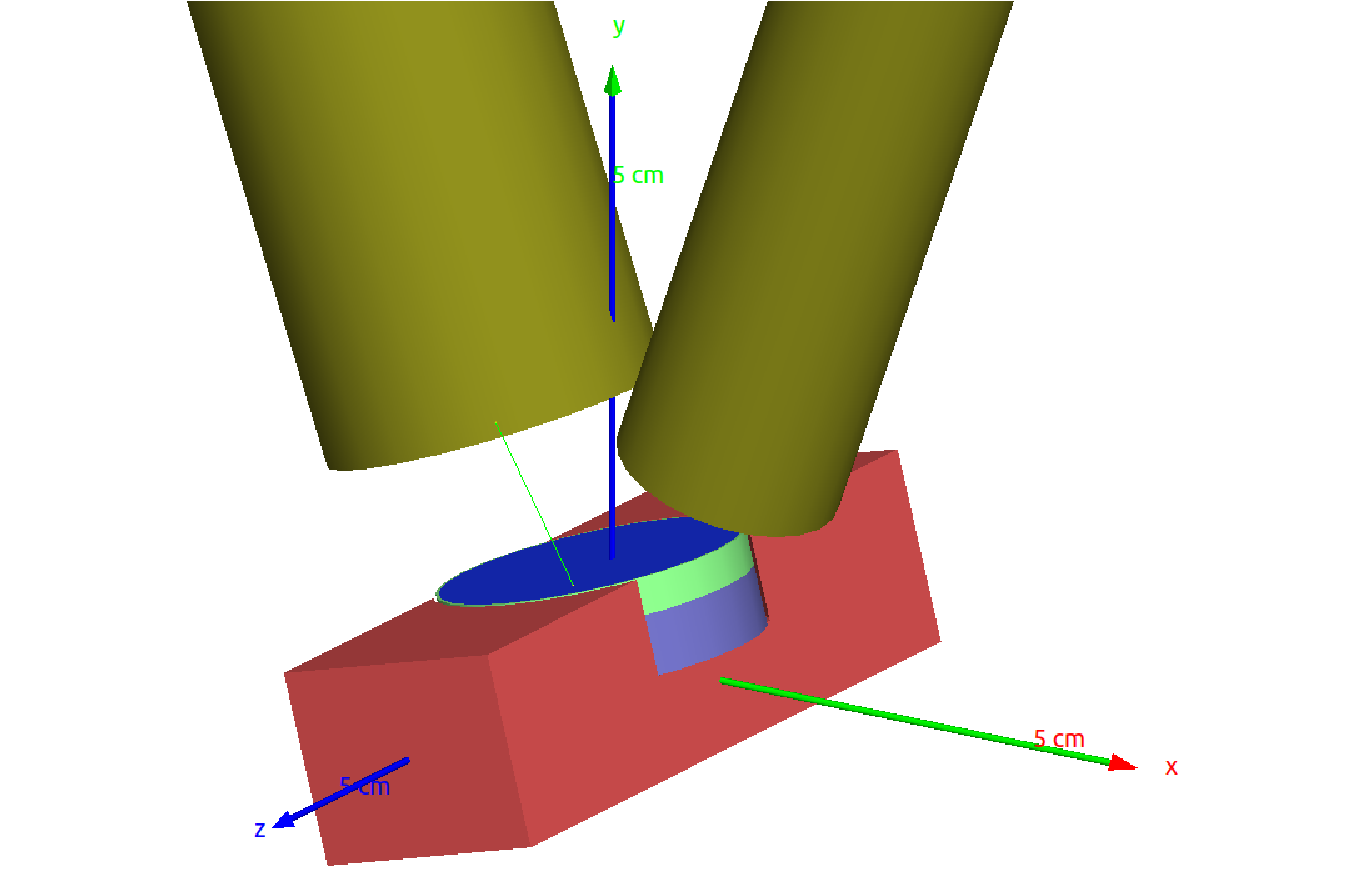}
\caption{\label{fig:photo_lum_setup} The Geant4 drawing depicts the photo-luminescence setup. The ThF$_4$ film is represented by the blue narrow disk, and the substrate (Si or ZnSe) is depicted by a light green disk. The sample is placed on the copper support (red rectangle) held by aluminium rings (violet disks). The two PMTs are illustrated as large yellow cylinders above the sample holder. A thin green line indicates a photon emitted from the sample and detected in the UV PMT.}
\end{center}
\end{figure}

The experimental setup for measuring photo-luminescence involves a VUV light source (described in section~\ref{sec:source}) and a detection system. The sample is first placed in the irradiation station where it is exposed to VUV light for approximately 5 minutes. After the exposure, the lamp is switched off and the sample is moved to the measurement station. The detection system 
comprises a solar-blind, VUV-sensitive, Hamamatsu PMT R6835 and a veto Hamamatsu R1450 PMT, which is sensitive to visible light (see Fig.~\ref{fig:photo_lum_setup}). 
The photo-luminescence induced by the VUV light during the irradiation is then measured for about 10 minutes. These measurements are repeated multiple times with different samples to minimize statistical uncertainties and eliminate various backgrounds.

\begin{figure}
\begin{center}
\includegraphics[scale=1.00]{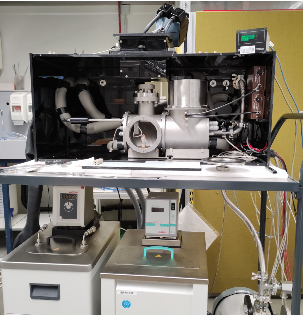}
\caption{\label{fig:photo_lum_setup_foto} Photo of the irradiation setup.}
\end{center}
\end{figure}

The stations for irradiation and measurement are kept under vacuum with a residual pressure of less than 5$\times$10$^{-4}$~mbar to prevent the absorption of VUV photons in air and minimize background luminescence from the atmospheric gas mixture. The vacuum is maintained continuously using a root pump and monitored by a pressure sensor. The macroscopic absorption cross-section of VUV light in O$_2$ or air is about 500~1/cm~\cite{uv_in_o2} under normal temperature and pressure conditions. Since the distance between the sample and the PMT is approximately 2~cm, at the residual pressure we anticipate a residual absorption of less than 0.05\%. Fig.~\ref{fig:photo_lum_setup_foto} shows a photo of the apparatus.

\begin{figure}
\begin{center}
\includegraphics[bb=4cm 2.5cm 20cm 26cm, angle=270, scale=0.35]{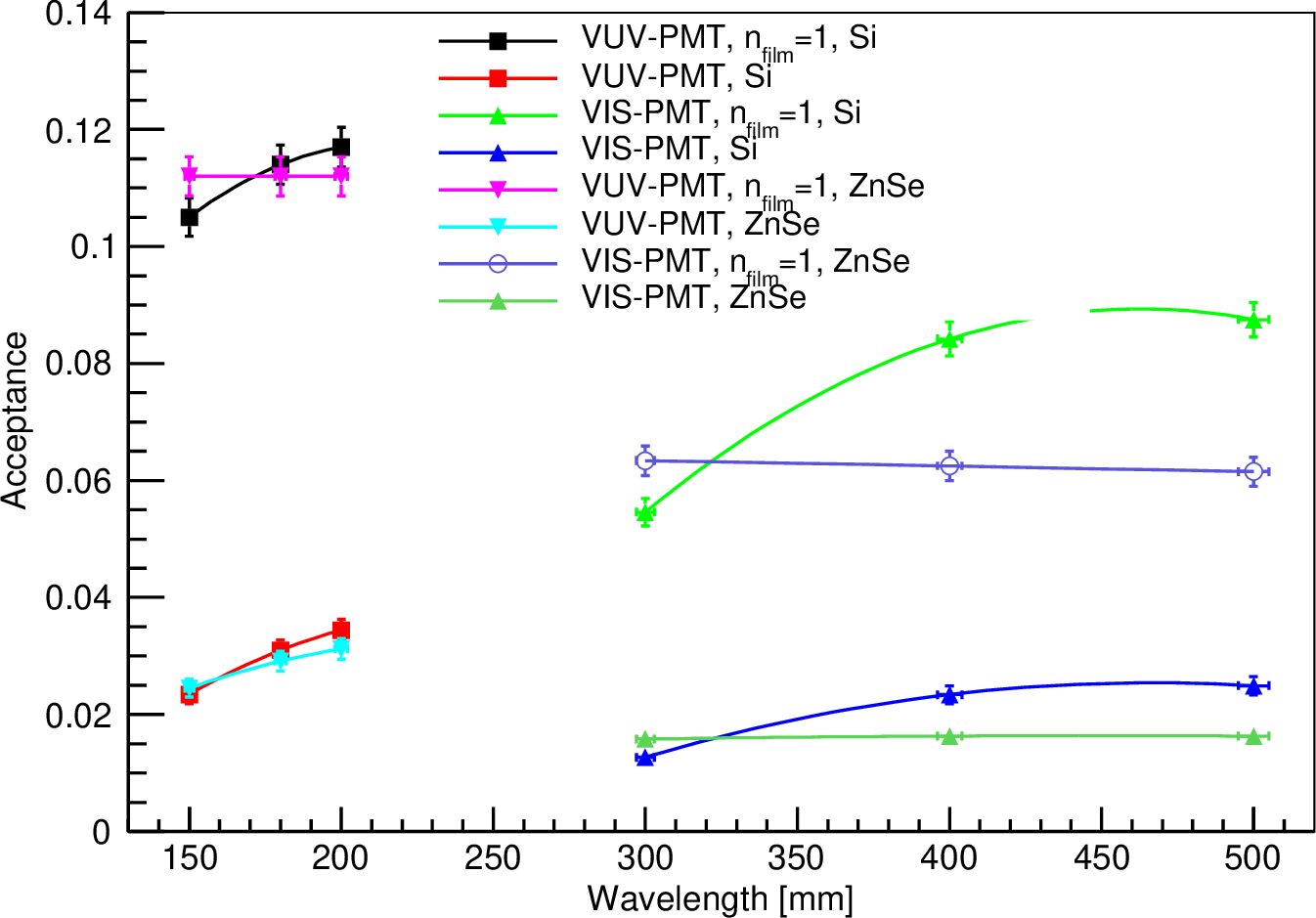}
\caption{\label{fig:photo_lum_acc} Acceptances of VUV and visible (VIS) PMTs for different wavelengths.
The acceptances indicated with $n_{film}=1$ are shown to provide an estimate of the TIR effect in the ThF$_4$ sample.}
\end{center}
\end{figure}

We used Monte Carlo simulations with the Geant4.11.2 library to estimate the acceptance of the two PMTs. The simulations considered the geometry and material properties of the sample holder, the sample, and the two PMTs. Photo-luminescence photons were generated uniformly and isotropically within the ThF$_4$ film. We then calculated the fraction of photons reaching each PMT photocathode area to estimate the acceptance. Fig.~\ref{fig:photo_lum_acc} displays the results for both PMTs. Since the apparatus lacks spectroscopic capabilities, we conducted acceptance simulations for a few wavelengths within the sensitivity range of each PMT. Additionally, to assess the effect of Total Internal Reflection (TIR) in the ThF$_4$ sample, we also provided results with $n_{film}=1$. 
TIR significantly reduced the acceptance, from approximately 11\% to 3\% for the VUV PMT, and from about 7\% to 2\% for the visible PMT. This reduction can be partially mitigated in thin films due to frustrated TIR, which allows about 50\% transmission at 1/3 of the wavelength. Therefore, for a 50-nm-thick ThF$_4$ film, we expected an acceptance of about 7\%.

The acceptance for ThF$_4$ on Si substrate slightly increased with wavelength due to a rise in silver-based mirror reflectivity. The ZnSe-based sample displayed almost constant acceptance distributions.

The measured luminescence spectra can be divided into two main wavelength ranges: 120 to 200~nm for the VUV PMT and 300 to 650~nm for the visible PMT. In the VUV range, the main contribution to the counting is likely from photons around 180~nm, where the quantum efficiency (QE) of the VUV PMT is about 1\%. For the visible PMT, the most likely wavelength is around 420~nm, which corresponds to the maximum QE point of about 25\%.


\section{Photo-luminescence measurements}\label{sec:photo_lum_exp}
\begin{figure}
\begin{center}
\includegraphics[scale=0.45]{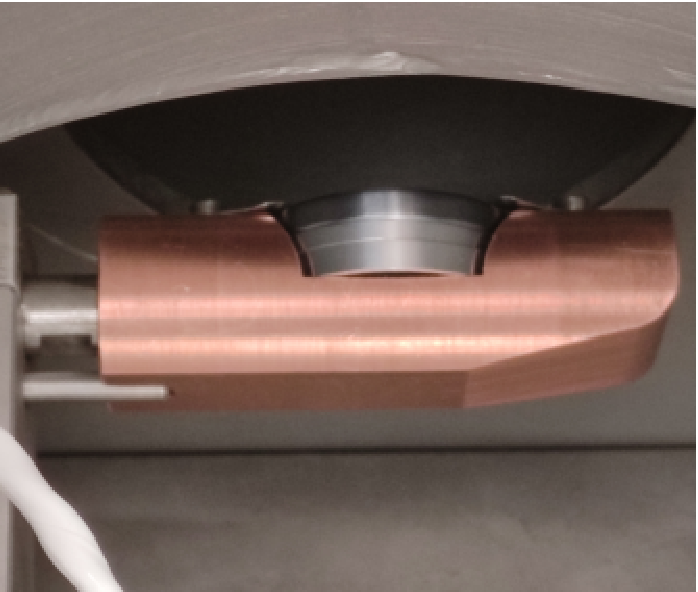}
\caption{\label{fig:sample_photolum_foto} Picture of the sample holder installed under the VUV-lamp.}
\end{center}
\end{figure}

To measure photo-luminescence, we mounted the samples (one at a time) on the movable samples holder as shown in Fig.~\ref{fig:sample_photolum_foto}. A vacuum of $5\times 10^{-4}$ mbar was created and maintained during irradiation and measurements using a root pump. We performed many successive measurements by irradiating the sample with the VUV lamp for 5 minutes (some runs were recorded after 15 minutes of irradiation) and moved the sample to the measurement station after turning off the lamp. At the measurement station, data acquisition started about 10 seconds after the VUV lamp shutdown.

\subsection{UV photo-luminescence measurements}\label{sec:uv_photo_lum_exp}
\begin{figure}
\begin{center}
\includegraphics[bb=4cm 2.5cm 20cm 26cm, angle=270, scale=0.27]{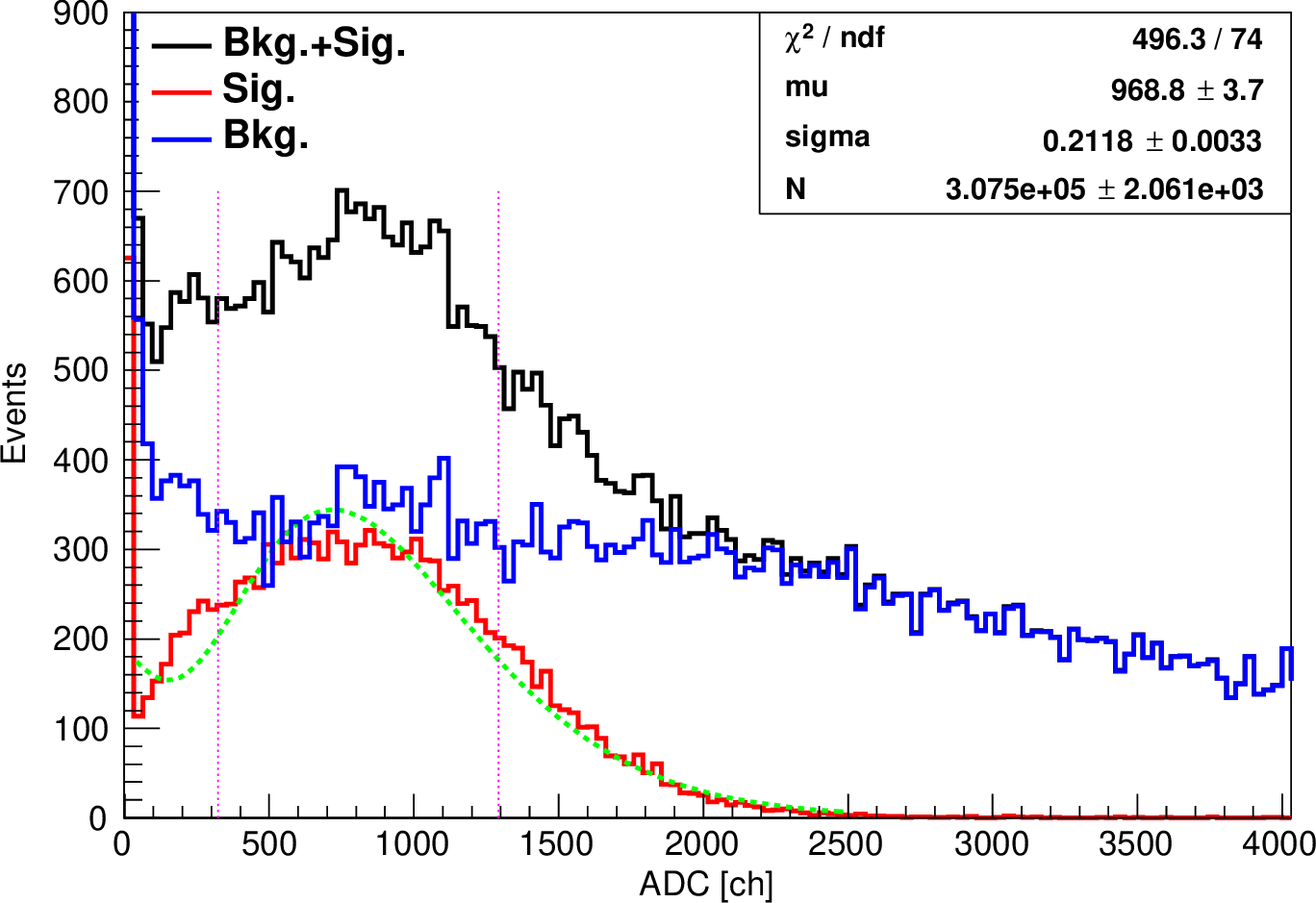}
\caption{\label{fig:uv_calib_noise} The Pulse Height Distribution (PHDs) of the VUV lamp signal measured by the UV PMT is represented by red histogram. The intrinsic background  is shown in the black histogram), and their difference is indicated by the blue histogram. A Polya fit to the signal data is depicted by the green dashed curve. Two vertical magenta dotted lines are used to show the selected ADC range, which maximizes the signal-to-background ratio while preserving 69\% of the signal statistics.}
\end{center}
\end{figure}

The expected signal rate is very low and is of a similar magnitude as the background. To make better use of the available experimental data, the definition of the signal was adjusted to enhance the signal-to-background ratio. To achieve this, the Pulse Height Distributions (PHDs) of the charge collected on the PMT anode were calibrated using the VUV lamp with the inter-chamber shutter lid slightly open. Fig.~\ref{fig:uv_calib_noise} shows the obtained PHD compared to the intrinsic background spectrum acquired with the sample removed. By comparing the two, we were able to determine the best  selected ADC interval from 1/3 to 4/3 of the Single Photo-Electron (SPE) amplitude, while still preserving 69\% of the signal statistics. This selection was applied to all the data discussed below and referred to as the SPE rate. A similar procedure was carried out for the visible PMT data.

\begin{figure}
\begin{center}
\includegraphics[bb=4cm 2.5cm 20cm 26cm, angle=270, scale=0.35]{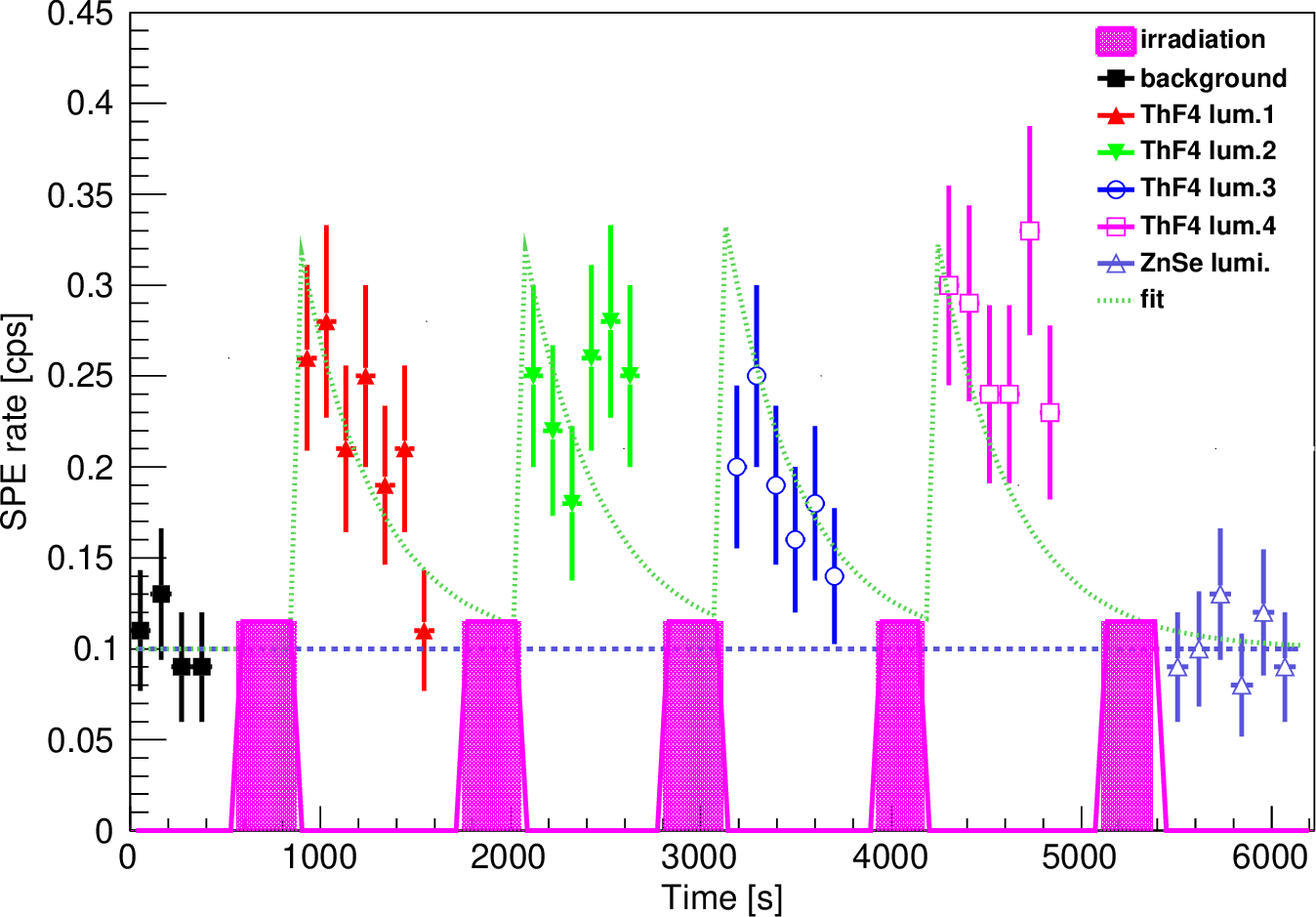}
\caption{\label{fig:uv_photo_lum_znse_delays} UV photo-luminescence of ThF$_4$+ZnSe in absolute time scale, including irradiation intervals between measurements. The dark-green dotted curve shows the expected photo-luminescence with a maximum amplitude of UV photo-luminescence of 0.23~Hz and a decay time of 400~s. The light-blue dashed line shows the average background level.}
\end{center}
\end{figure}

During the night before the photo-luminescence measurements, the UV-PMT background was 0.1~Hz.
The background measurement was repeated just before the irradiation of the sample, as shown in the time evolution of the UV-PMT SPE rate reported in Fig~\ref{fig:uv_photo_lum_znse_delays}. The sequence of ThF$_4$+ZnSe sample irradiations and photo-luminescence decay measurements roughly follows the pattern indicated by the fitted curve. The UV photo-luminescence starts at about 0.23~cps and decays with a lifetime of roughly 400~s. The last measurement in Fig~\ref{fig:uv_photo_lum_znse_delays} represents the photo-luminescence of the ZnSe side of the target, with the sample flipped upside-down. This observed rate aligns with the previously measured background, indicating that ZnSe is opaque to UV light and does not produce significant contamination in UV photo-luminescence.

The low event rate affects the statistical precision of the data, resulting  in relatively uncertain estimates for both the photo-luminescence amplitude and the lifetime. This uncertainty allows for the neglect of systematic uncertainties related to the stability of background yield. The difference between different runs is estimated to be of the same magnitude as the statistical uncertainties.

\begin{figure}
\begin{center}
\includegraphics[bb=4cm 2.5cm 20cm 26cm, angle=270, scale=0.35]{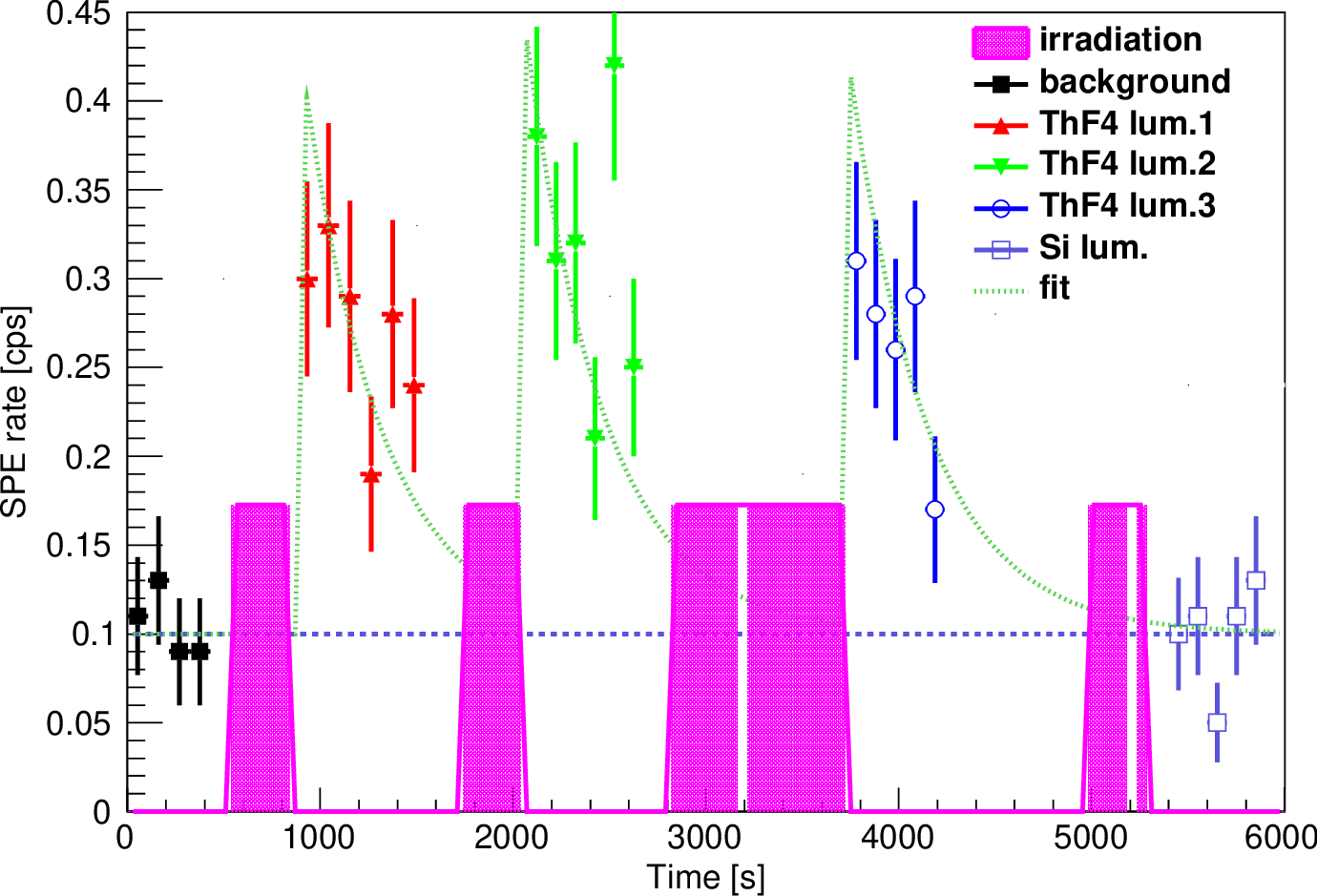}
\caption{\label{fig:uv_photo_lum_si_delays} UV photo-luminescence of ThF$_4$+Si in absolute time scale, including irradiation intervals between measurements. The dark-green dotted curve shows the expected photo-luminescence with a maximum amplitude of UV photo-luminescence of 0.23~Hz and a decay time of 400~s. The light-blue dashed line shows the average background level.}
\end{center}
\end{figure}

The irradiated ThF$_4$+Si sample showed an expectedly higher UV photo-luminescence, peaked at about 0.35~Hz, as shown in Fig~\ref{fig:uv_photo_lum_si_delays}. This outcome was anticipated due to the increased thickness of ThF$_4$ on the Si substrate, which changed from 200~nm to 300~nm. This change implied  an enhancement of 3/2, which was effectively observed. These data are also compatible with a UV photo-luminescence lifetime of 400~s. In the last measurement of the ThF$_4$ side, the irradiation time was extended from 5 to 15 minutes. However, the observed UV photo-luminescence remained consistent with previous measurements. This suggests that the rate of UV photo-luminescence is not influenced by the duration of irradiation, indicating that the excitation of fluorescent levels reaches saturation in a time frame shorter than the chosen 5~min.

Summarizing, in the UV range, we observed a photo-luminescence of 0.23~Hz/0.69 from a 200~nm thick ThF$_4$ film, which decayed in about 400~s. The ThF$_4$ film had a surface area of 4.84~cm$^2$ whilst the UV PMT acceptance was about 0.03. Therefore, the estimated total UV photo-luminescence yield is 11~photons/s/PMT quantum efficiency, which varies from 0.1 to 0.001 in the range from 150 to 190~nm.

\subsection{Visible photo-luminescence measurements}\label{sec:vis_photo_lum_exp}
The measurements of photo-luminescence in the visible range were performed at the same time as the UV measurements. The visible-range PMT had a similar acceptance, but much higher quantum efficiency in its sensitivity interval. Moreover, the ZnSe sample is transparent to visible light, which creates a background source.

\begin{figure}
\begin{center}
\includegraphics[bb=4cm 2.5cm 20cm 26cm, angle=270, scale=0.35]{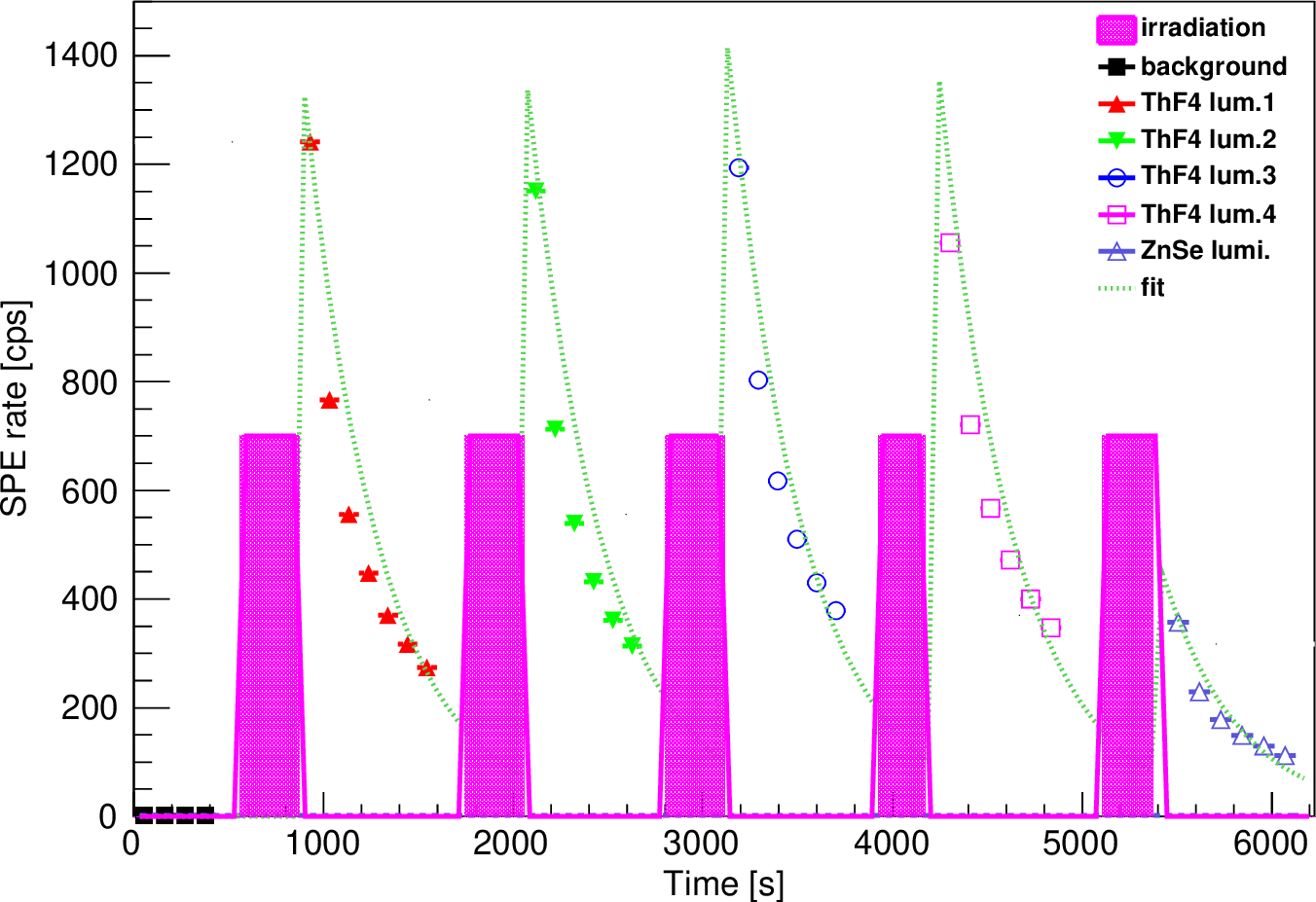}
\caption{\label{fig:vis_photo_lum_znse_delays} Visible photo-luminescence of ThF$_4$+ZnSe in absolute time scale including irradiation intervals between measurements. The dark-green dotted line shows the expected photo-luminescence with a maximum amplitude of visible photo-luminescence of 1.4~kHz and a decay time of 400~s.}
\end{center}
\end{figure}

The analysis followed the same method applied to the UV photo-luminescence data. For the visible PMT,, the observed background SPE rate was approximately 2 Hz, which is negligible compared to the observed photo-luminescence shown in Fig.~\ref{fig:vis_photo_lum_znse_delays}. The ThF$_4$+ZnSe sample, irradiated for 5~minutes, exhibited a peak photo-luminescence of about 1.4~kHz. The time distribution of the photo-luminescence cannot be precisely described by a single exponential. However, a basic fit to the data gives a visible photo-luminescence lifetime of about 400~s, consistent with UV range measurements. Data obtained after ZnSe irradiation (ThF$_4$ not exposed to the UV lamp and PMT) showed approximately 30\% contribution of substrate photo-luminescence in the main observable.

The irradiated ThF$_4$+Si sample showed a similar visible photo-luminescence of about 1~kHz and a similar lifetime. However, due to some contamination, a continuous luminescence build-up was observed independently from the target material and we excluded these data from the analysis.

Summarizing, we have observed a 1.4~kHz$\times$0.70/0.70 photo-luminescence in the visible range from a 200~nm thick ThF$_4$ film, decaying in about 400~s. The ThF$_4$ film had a surface area of 4.84~cm$^2$ whilst the visible PMT acceptance was about 0.015. Therefore, the total visible photo-luminescence yield can be estimated as 4.7$\times$10$^5$~photons/s.


%
\section{Radio-luminescence setup and Monte Carlo simulations}\label{sec:radio_mc_sim}
\begin{figure}
\begin{center}
\includegraphics[scale=0.20]{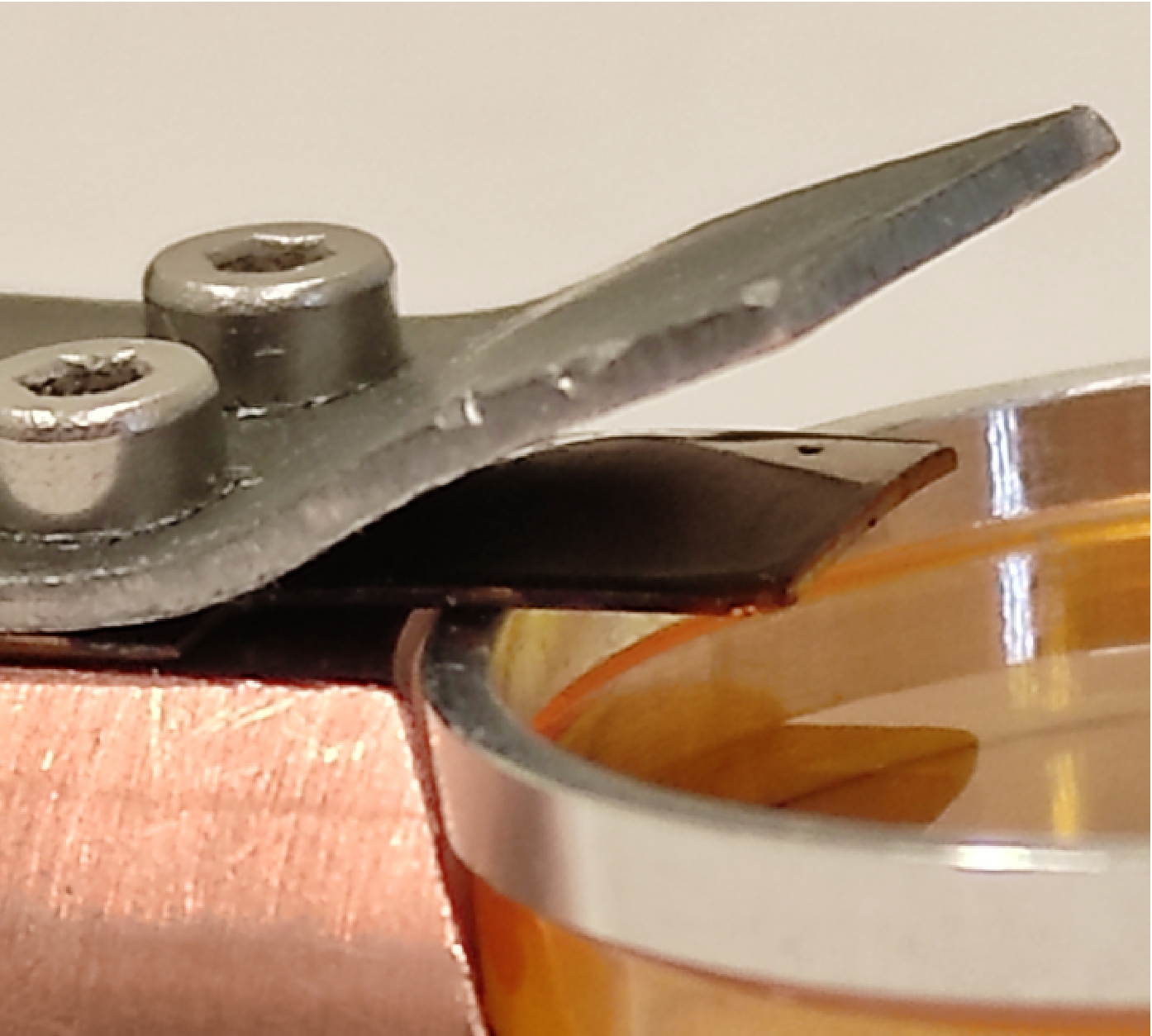}
\caption{\label{fig:radio_lum_setup_foto} The $^{241}$Am $\alpha$-source was positioned 2~mm above the ZnSe sample. To illustrate the position of the $\alpha$-source substrate, the lead sheet was bent while capturing the photo.}
\end{center}
\end{figure}

The experimental setup for the radio-luminescence measurements uses of a $^{241}$Am $\alpha$-source and the detection system described in section~\ref{sec:photo_mc_sim}. The $^{241}$Am $\alpha$-source was positioned 2~mm above the sample, as shown in Fig.~\ref{fig:radio_lum_setup_foto}. The active surface of the source faced downwards, in the opposite direction of the PMTs, to prevent direct $\alpha$-particle hits. Additionally, a 1~mm-thick lead sheet covered the source substrate to absorb 60~keV gamma and other X-rays traveling towards the PMTs. The sample was initially placed in the irradiation station to begin background data acquisition. Subsequently, the sample was moved to the measurement station. The geometry was designed such that the PMTs only detected the secondary emission from the sample, enabling measurement of the radio-luminescence induced by the $\alpha$ source over a period of approximately 10 minutes. These measurements were repeated multiple times to minimize statistical uncertainties and were conducted using different samples to eliminate various backgrounds.

We conducted Monte Carlo simulations using the Geant4.11.2~\cite{geant4} library, as outlined in section ~\ref{sec:photo_mc_sim}. The simulations involved adding an $\alpha$-source and its shielding, as well as the ThF$_4$ scintillation as an additional light production mechanism.  The $\alpha$ particles were uniformly and isotropically generated on the active surface of the 2~mm diameter source substrate. We found that a 0.264 fraction of $\alpha$ particles reached the ThF$_4$ film. Since the light yield and spectrum of ThF$_4$ scintillation were unknown, we used scintillation parameters adapted to the measurement results. Given that the $\alpha$ particle energy was fixed at about 5.5~MeV and the mean energy loss in ThF$_4$ layers was 84 and 126~keV for 200 and 300~nm films, light quenching was excluded and we used absolute light yields separately for the VUV and visible ranges.

\begin{figure}
\begin{center}
\includegraphics[bb=4cm 2.5cm 20cm 26cm, angle=270, scale=0.35]{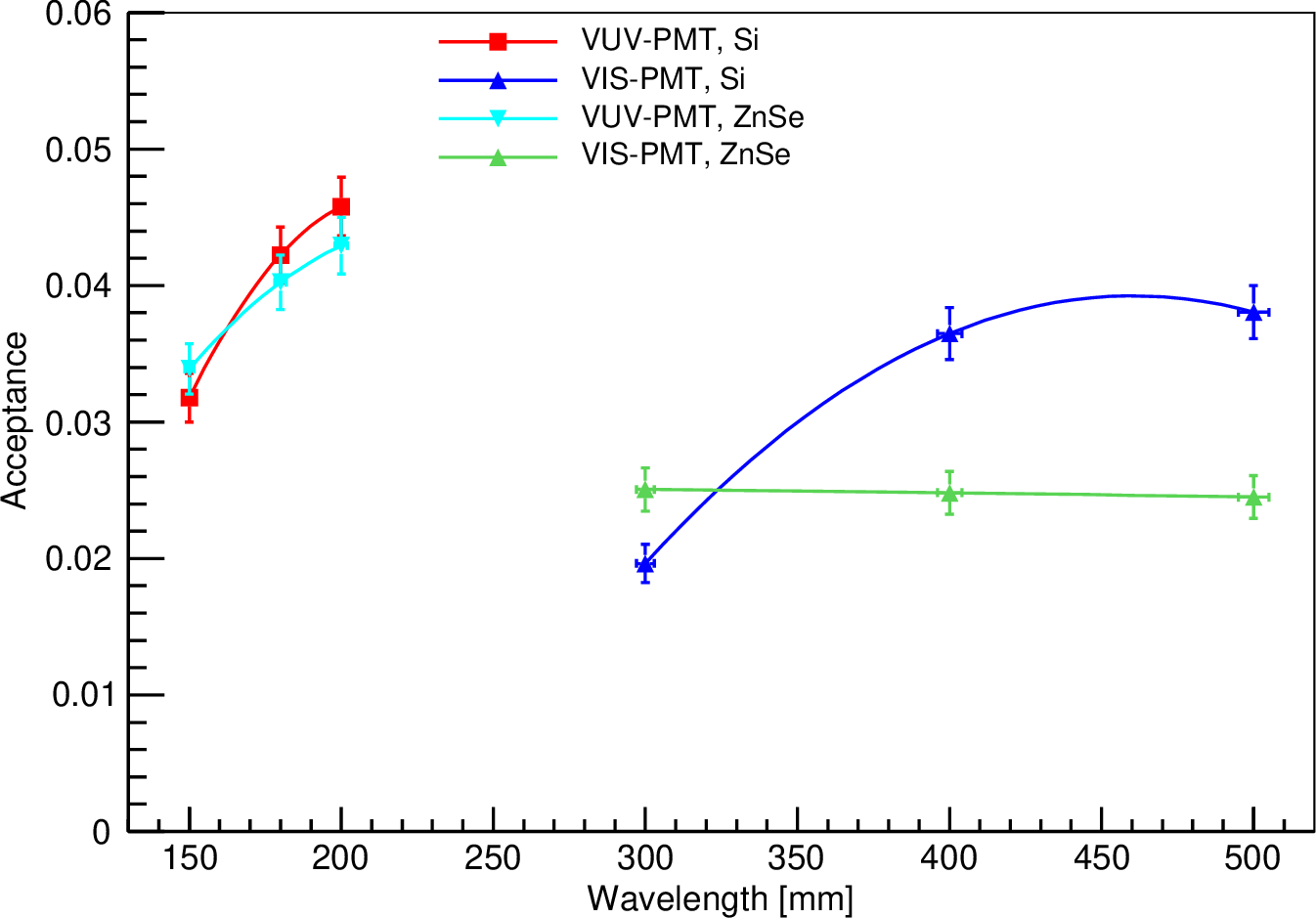}
\caption{\label{fig:radio_lum_acc} Acceptances of the VUV and visible (VIS) PMTs for scintillation photons generated by radio-luminescence in the ThF$_4$ film as a function of their wavelength.}
\end{center}
\end{figure}

We calculated the percentage of scintillation photons that reach the PMT photocathode area to estimate the acceptance. Figure \ref{fig:radio_lum_acc} shows the results for each PMT.

Because the equipment cannot perform spectroscopic analysis, we conducted acceptance simulations for several wavelengths within the sensitivity range of each PMT. The acceptance for ThF$_4$ on Si substrate slightly increases with wavelength due to the rising reflectivity of the silver-based mirror. The ZnSe-based sample exhibits an almost constant acceptance distribution.

\begin{figure}
\begin{center}
\includegraphics[scale=0.50]{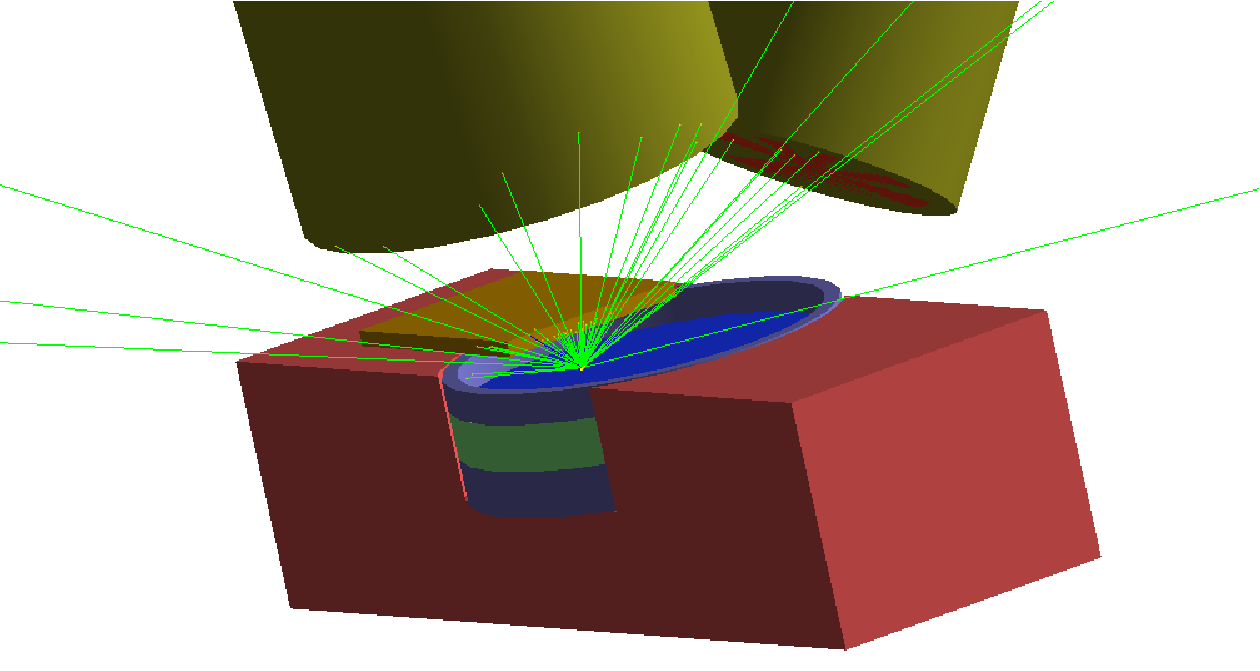}
\caption{\label{fig:radio_lum_setup} In this Geant4 drawing of the radio-luminescence setup, the source substrate is depicted by a bronze plate.
The ThF$_4$ film is represented by the blue narrow disk, and the substrate (Si or ZnSe) is shown as a light-green disk mounted on the copper support (red rectangle) and sustained by aluminium rings (violet disks). The two PMTs are illustrated as large yellow cylinders above the sample holder. The thin green lines in the drawing represent scintillation photons emitted from the sample and reaching the PMTs.}
\end{center}
\end{figure}

 To simulate scintillation, common parameters of scintillator materials were chosen, namely 0.5 photons/keV with Birks coefficient set to zero and scintillation decay time set to 1~ns.

\section{Radio-luminescence measurements}\label{sec:radio_lum_exp}
Measurements were taken at room temperature, with a chamber pressure of 5$\times$10$^{-4}$~mbar, using a specially prepared $^{241}$Am $\alpha$ source of approximately 1.4 kBq activity (+-10\% uncertainty). The source was positioned about 2~mm above the commercial sample and was covered by a 1~mm thick lead sheet, which was 2 mm wider on each side, to shield the PMTs from the direct background of 60~keV $\gamma$ rays. The entire setup, including the source and shielding, covered nearly half of the sample, as shown in Fig.~\ref{fig:radio_lum_setup_foto}.

\subsection{UV radio-luminescence}\label{sec:uv_radio_lum_exp}
\begin{figure}
\begin{center}
\includegraphics[bb=4cm 2.5cm 20cm 26cm, angle=270, scale=0.27]{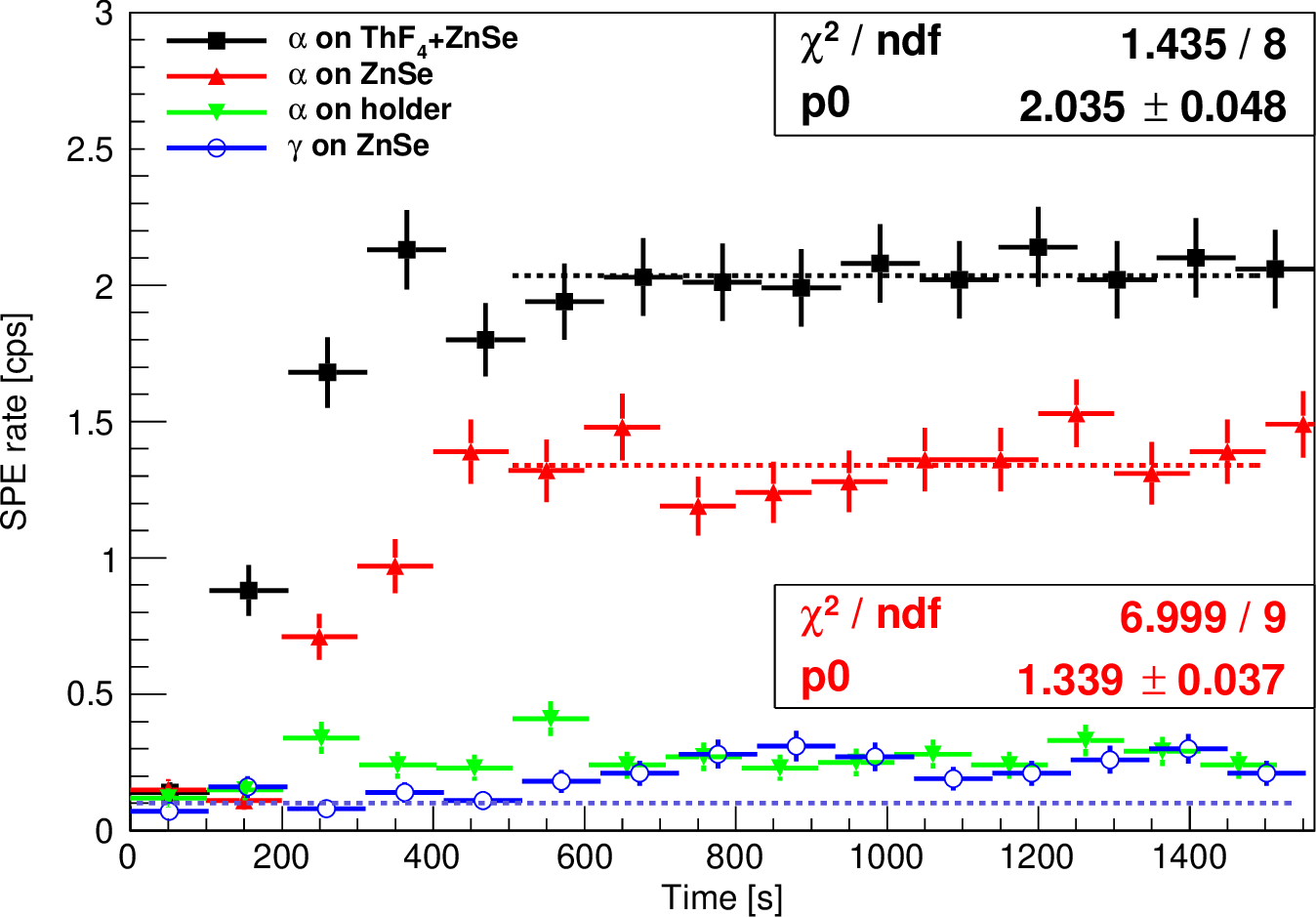}~%
\includegraphics[bb=4cm 2.5cm 20cm 26cm, angle=270, scale=0.27]{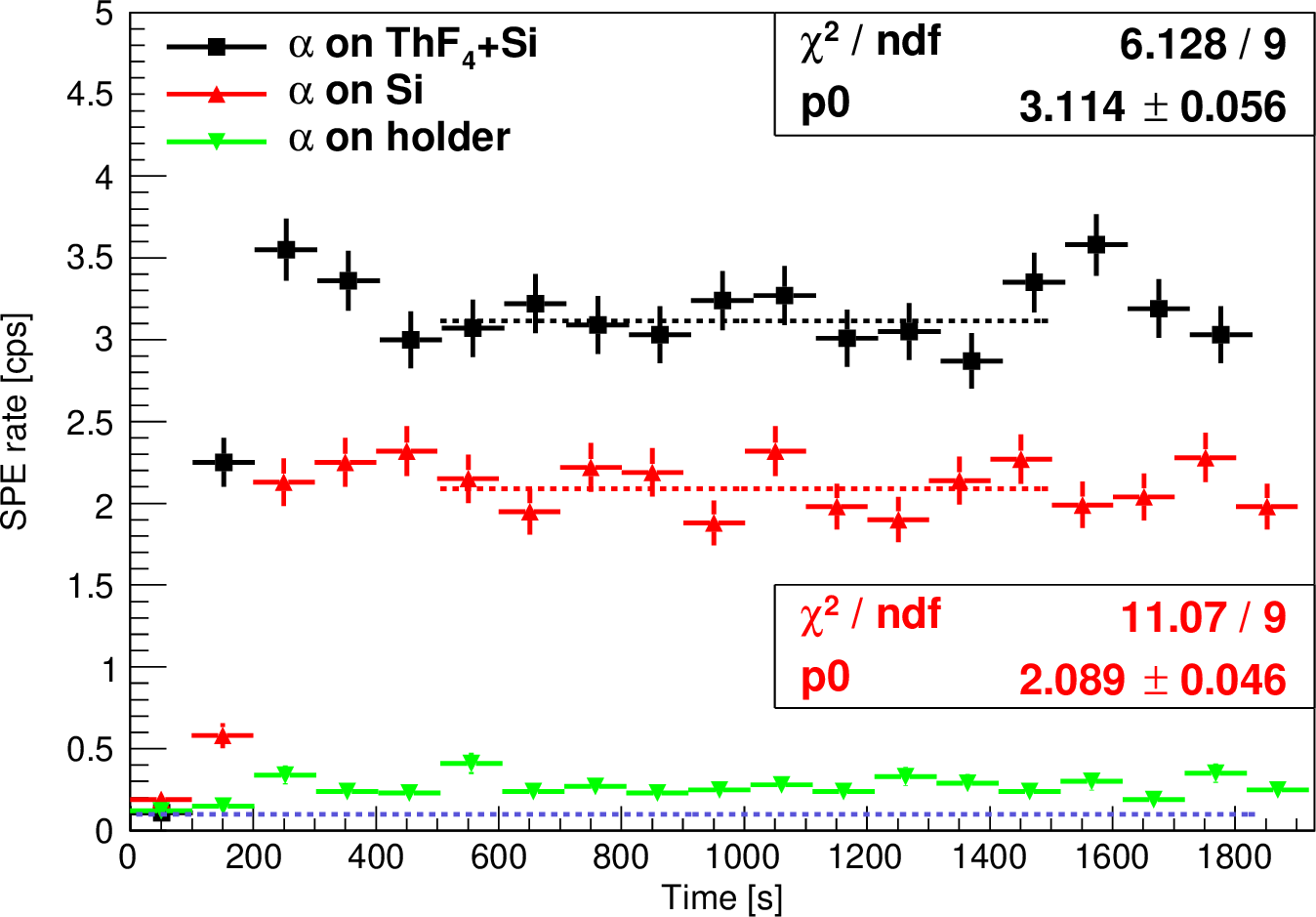}
\caption{\label{fig:uv_radio_lum_both} UV radio-luminescence of ThF$_4$+ZnSe (left) and ThF$_4$+Si (right) samples. The plots compare the measured event rates on both sides of each sample, the measurement with $\alpha$-source wrapped into a 100~$\mu$m foil to suppress $\alpha$-particles, and the measurements without sample. The small rates in the first 200~s are measured with the sample retracted into the irradiation chamber for the background check.}
\end{center}
\end{figure}

In the UV range, all signal rates are relatively low but emerge above the background of 0.1~Hz, as shown in Fig.~\ref{fig:uv_radio_lum_both}. To monitor the PMT intrinsic background, all measurements started with the sample retracted into the irradiation station. After about 200~s (700~s for the $\gamma$-background measurement), the sample was transferred into the measurement station. This explains the rapid rise of the radio-luminescence rates at the beginning of the measurements. The steady-state radio-luminescence was obtained by averaging the rates starting from 500~s after the beginning of the run. Since the 5~MeV $\alpha$-particle crosses the 200-300~nm film, the radio-luminescence rate was estimated from the difference of measurements on opposite sides of the sample. In particular, for the 200~nm-tick ThF$_4$ deposited on ZnSe, the ThF$_4$ radio-luminescence was obtained from the difference:
2.04-1.34=0.70$\pm$0.06~Hz.
Instead, for the 300~nm-thick ThF$_4$ deposited on Si, the same difference yields:
3.11-2.09=1.03$\pm$0.07~Hz.
The results are consistent with each other: the Si-mirror contains 1.5 times more ThF$_4$, thus the observed UV-ratio Si/ZnSe, 
1.03/0.70=1.46$\pm$0.16,
is in perfect agreement with the ratio of thicknesses.
Also in this case, the statistical uncertainties of about 7-8\% are dominant, allowing for the neglect of systematic uncertainties caused by the intrinsic instability of the background rate.

The $^{241}$Am source also emits $\gamma$-rays, with a 35.9\% branching fraction for the 60 keV emission. ZnSe is a good scintillator, so the impact of 60~keV $\gamma$ rays on ZnSe will produce scintillation light. To measure this effect, we wrapped the $^{241}$Am source into a 100~$\mu$m thick Aluminium foil. The foil absorbed all the $\alpha$ particles emitted by the source, while the $\gamma$ rays were unaffected. This contribution is shown by the blue circles in Fig.~\ref{fig:uv_radio_lum_both}. Since ZnSe blocks UV light but allows $\gamma$ rays to penetrate and interact within the material, we did not observe a significant contribution from the $\gamma$ rays on ZnSe in UV radio-luminescence.

The measured radio-luminescence event rates must be normalized to the known $^{241}$Am alpha source activity (1.4 kBq), accounting for the probability for an $\alpha$-particle to reach the ThF$_4$ layer (0.264), as given in section~\ref{sec:radio_mc_sim}. Therefore, the observed (1.03$\pm$0.07~Hz)/0.69 of UV radio-luminescence must be divided by 370~Hz $\alpha$-particle impacts on the 300~nm ThF$_4$ film. This provides an estimate of the UV light detection probability per each $\alpha$-particle interaction of about 0.0040$\pm$0.0003. The deduced probability represents the convolution of the PMT acceptance (0.02-0.04), PMT quantum efficiency, and ThF$_4$ scintillation spectrum. Thus, one can estimate the ThF$_4$ UV light yield $Y_{UV}(\lambda)$ as follows:
\begin{equation}\label{eq:light_yield}
Y_{UV}(\lambda) = \frac{0.0040}{QE_{UV}(\lambda) \epsilon_{UV}(\lambda)} ~,
\end{equation}
\noindent where $QE_{UV}(\lambda)$ is the UV PMT quantum efficiency and $\epsilon_{UV}(\lambda)$ the acceptance of the UV PMT. For example, if we assume that the main contribution to the UV PMT counting comes from $\lambda$=180~nm photons, where $QE_{UV}(\lambda)$ of the PMT is still about 1\%, we obtain a ThF$_4$ light yield of about 0.09 photons/keV. In the case of $\lambda$=200~nm photons, the light yield could be as large as 15 photons/keV.


Given the small magnitude of the detection probability, the detection of double photons from the same alpha scintillation is unlikely. This was confirmed by the observation of the measured collected charge spectrum in the PMT, where the mean number of photo-electrons could not exceed a few per cent.

\subsection{Visible radio-luminescence measurements}\label{sec:vis_radio_lum_exp}
\begin{figure}
\begin{center}
\includegraphics[bb=4cm 2.5cm 20cm 26cm, angle=270, scale=0.27]{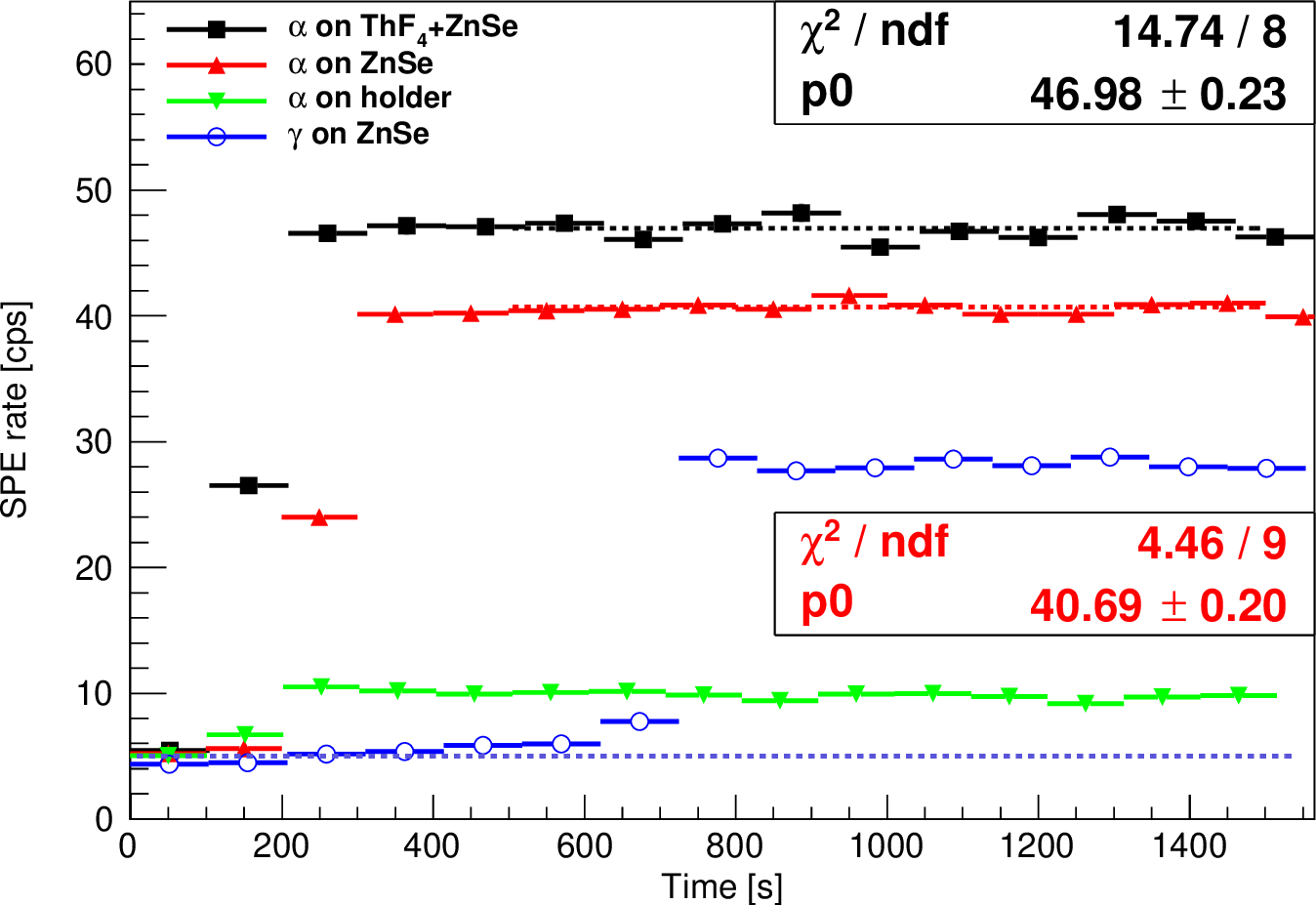}~%
\includegraphics[bb=4cm 2.5cm 20cm 26cm, angle=270, scale=0.27]{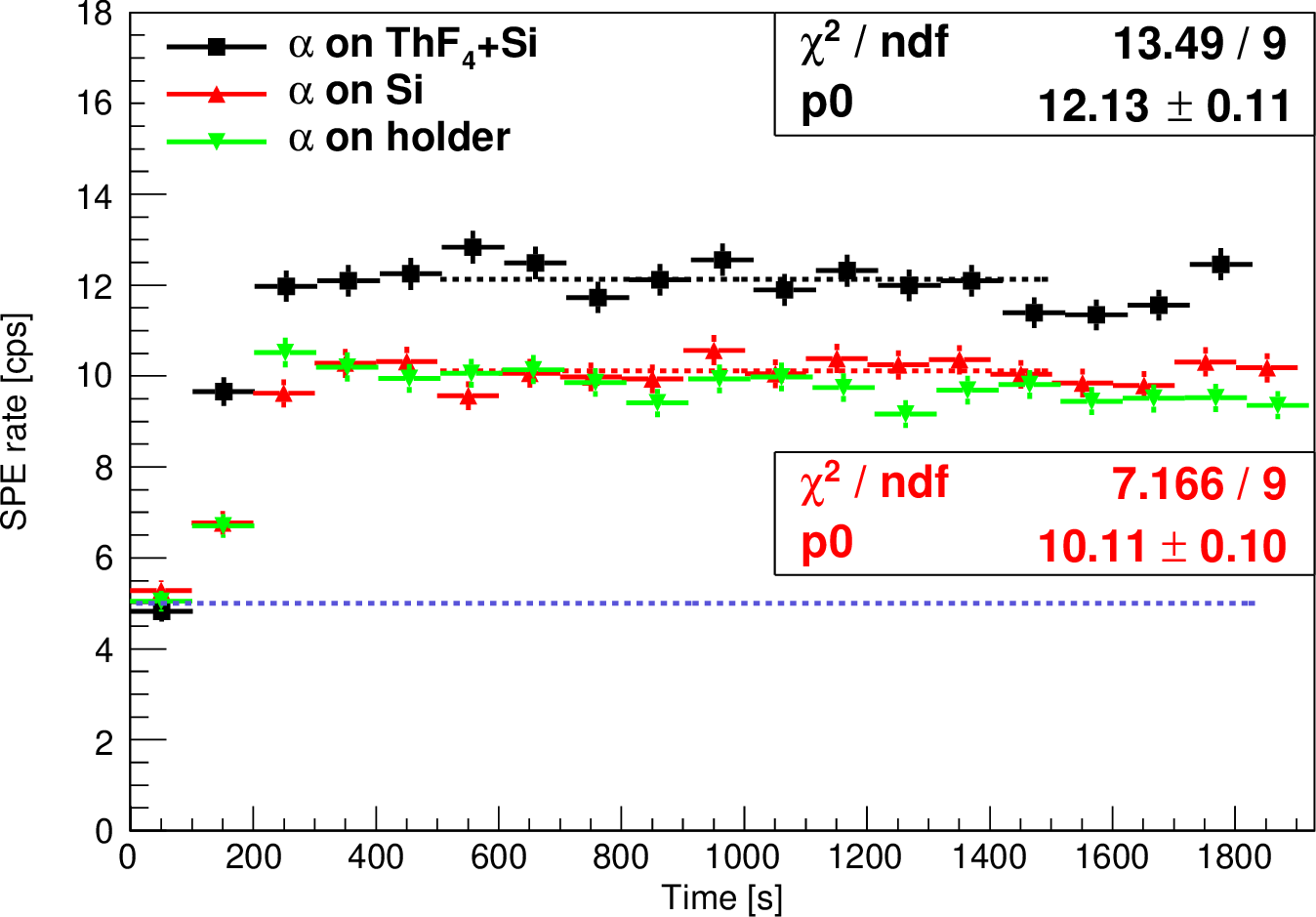}
\caption{\label{fig:vis_radio_lum_both} Visible range radio-luminescence of ThF$_4$+ZnSe (left) and ThF$_4$+Si (right) samples. The plots compare the measured event rates on both sides of each sample, the measurement with $\alpha$-source wrapped into a 100~$\mu$m foil to suppress $\alpha$-particles, and the measurements without sample. The small rates in the first 200~s are measured with the sample retracted into the irradiation chamber for the background check.}
\end{center}
\end{figure}

In the visible range, the signal rates are higher than in the UV, especially for the ZnSe sample, and significantly exceed the background of 5~Hz, as one can see in Fig.~\ref{fig:vis_radio_lum_both}. However, in the case of substrate radio-luminescence dominance (ZnSe), the optical transmission through the ThF$_4$ film must be considered.
For the complex coupling ZnSe$\to$ThF$_4$$\to$Air, the Fresnel transmittance of the light emitted at small angles to the sample normal is 1.13 times larger than the transmittance of the direct coupling ZnSe$\to$Air. Thus, the signal from ThF$_4$ on ZnSe was obtained from the difference of measurements on different sample sides, correcting for the Fresnel transmittance ratio:
47.0-40.7*1.13=1.0$\pm$0.3~Hz.
It is also worth noting that ZnSe is a good scintillator in the visible range, but opaque in UV. This is because the radio-luminescence from ZnSe alone dominates the measurements by almost an order of magnitude.

In the case of 300~nm of ThF$_4$ on Si substrate, the same difference (not requiring Fresnel correction since the Si substrate is opaque) yields
12.1-10.1=2.0$\pm$0.15~Hz.
In this case, the measurements performed on the Si substrate and sample holder alone agree, within statistical uncertainties. This suggests that the Si substrate does not emit visible light and the main background comes from the $\gamma$ interaction in the PMT or supporting materials. The latter background is the same for the two samples and accounts for about 12~Hz.


The results are again compatible with each other, as the Si-mirror has 1.5 times more ThF$_4$, thus the observed visible-ratio Si/ZnSe equals to
2.0/1.0=2.0$\pm$0.6
and agrees with the ratio of thicknesses within the uncertainties.

The observed 2~Hz of visible radio-luminescence from 370~Hz alpha particle impacts on ThF$_4$ film provides an estimate of 0.008 for the visible light detection probability per each $\alpha$-particle interaction. In this case, the wavelength dependence of the acceptance is relatively flat. Thus, it is safe to assume that the dominant contribution to visible PMT counting comes from 400~nm photons, where the QE of the PMT has the main peak. Therefore, in the visible range, according to Eq.~\ref{eq:light_yield}, the light yield of ThF$_4$ is approximately 0.007 photons/keV.


\section{Projection on signal rate expectations}\label{sec:projection_expect}
To enable a qualitative evaluation of the experiment, the results described above must be compared to the expected signal. Such a comparison will allow us to verify if the obtained background rejection is sufficient to identify the $^{229m}$Th decays. Therefore, we report here an estimate of the signal rate based on the available knowledge of $^{229m}$Th.

Resonant cross sections are generally described by the (non-relativistic) Breit-Wigner formula~\cite{PDG}:
\begin{equation}
\sigma(E) = \frac{2J+1}{(2S_1+1)(2S_2+1)} \frac{4\pi}{k^2}
\Biggl [ \frac{\Gamma^2/4}{(E-E_0)^2+\Gamma^2/4} \Biggr ]
B_{in} B_{out} ~,
\end{equation}
\noindent where $E$ is the center of mass  energy, $J$ is the spin of the resonance, and the number of polarization states of the two incident particles are $2S_1+1$ and $2S_2+1$. The center of mass  momentum in the initial state is k, $E_0$ is the center of mass energy at the resonance, and $\Gamma$ is the full width at half the maximum height of the resonance. The branching fraction for the resonance into the initial-state channel is $B_{in}$ and into the final-state channel is $B_{out}$.

\begin{figure}[!ht]
\begin{center}
\includegraphics[bb=1cm 2.5cm 20cm 16cm, scale=0.25, angle=270]{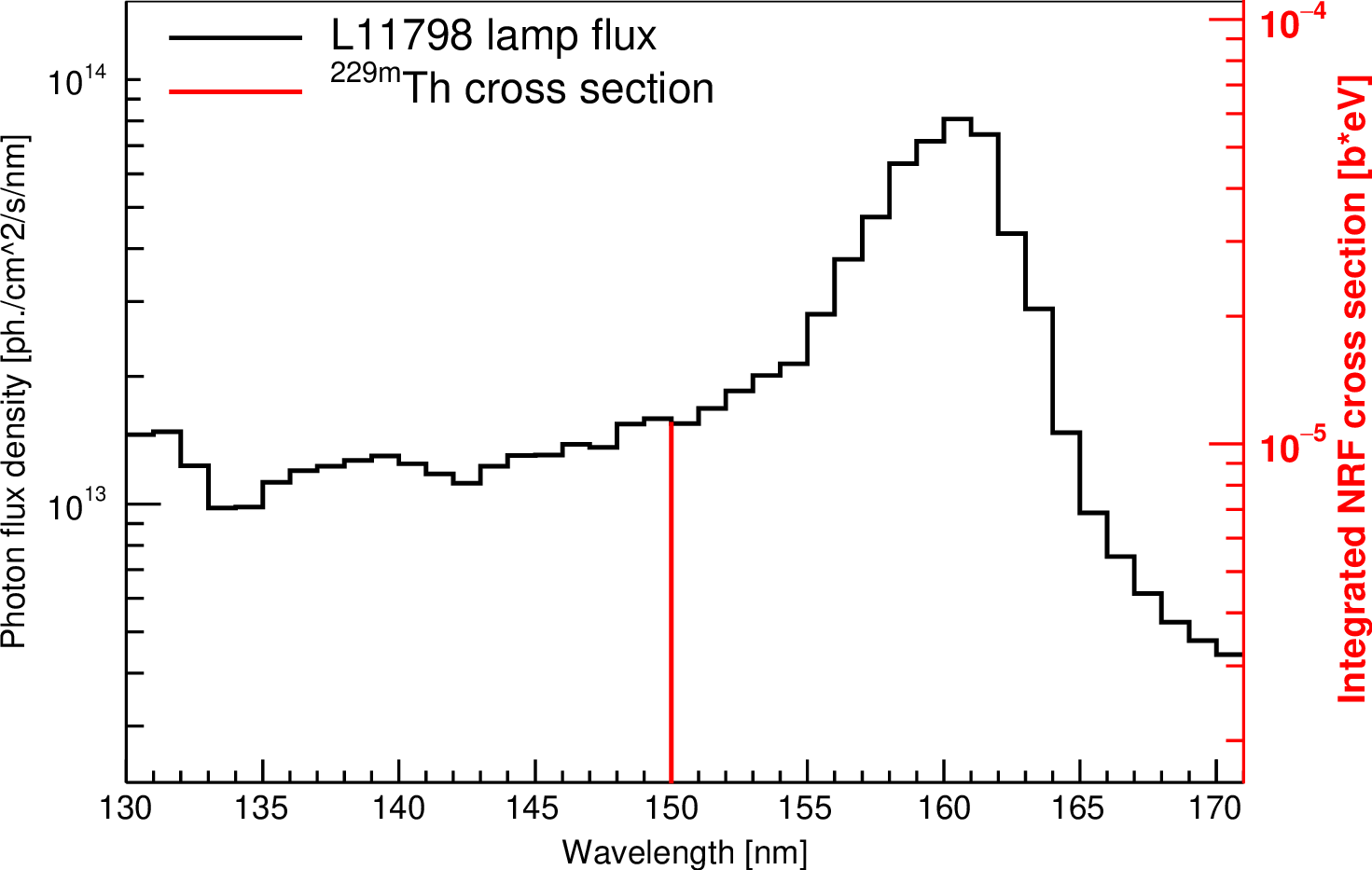}
\caption{\label{fig:source_flux_th229m} Photon flux at the sample location in the region of the $^{229m}$Th energy measured in section~\ref{sec:source}. The red line shows the cross-section peak of the $^{229m}$Th excitation (the peak width is much smaller than the resolution of the figure).}
\end{center}
\end{figure}

For a narrow resonance $\Gamma<<\Delta E$ (with $\Delta E$ being the energy resolution of the measurement), the peak can be approximated by a $\delta$-function. Moreover, in the case of the resonant photo-absorption, we have $B_{out}=1$, $k=2\pi/\lambda$ (here we use the system of units $\hbar = c = 1$). Finally, assuming that the IC decay branching is completely suppressed in ThF$_4$ ($B_{\gamma}=1$, $\Gamma_\gamma=\Gamma$) and defining $g=\frac{2J+1}{(2S_1+1)(2S_2+1)}$ we can rewrite the cross section as following:
\begin{equation}
\sigma(E) = g \lambda^2 \frac{\Gamma_\gamma}{2} \delta(E-E_0) ~.
\end{equation}
To obtain an estimate of the $^{229m}$Th excitation rate, the cross-section has to be convoluted with the incident photon flux. The absolute photon source flux $\frac{dN_\gamma}{dt d\lambda}(\lambda)$ of our VUV-lamp in the region of the $^{229m}$Th resonance is shown in Fig.~\ref{fig:source_flux_th229m}.

\begin{figure}[!ht]
\begin{center}
\includegraphics[bb=1cm 2.5cm 20cm 16cm, scale=0.25, angle=270]{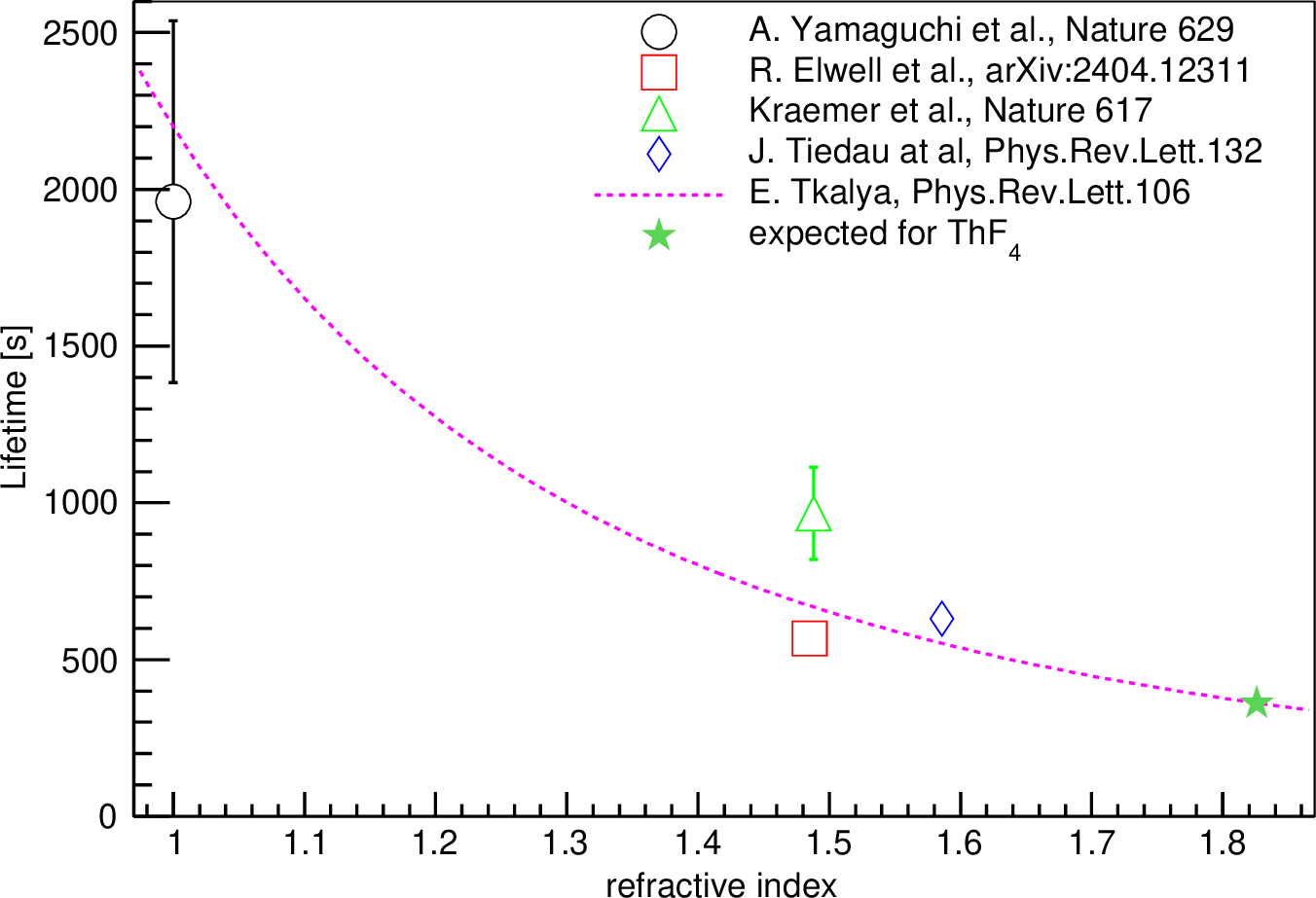}
\caption{\label{fig:lifetime_th229m} Lifetime of $^{229m}$Th measured in different media~\cite{Tiedau_tau_th229m_caf2,Kraemer_tau_th229m_mgf2,Elwell_tau_th229m_licaf,Yamaguchi_tau_th229m_vacuum} in comparison with theoretical prediction from Ref.\cite{Tkalya_gamma}.}
\end{center}
\end{figure}

In the narrow width of the $^{229m}$Th resonance, the photon flux can be considered constant at about 1.6 $\times$ 10$^{13}$ ph/(cm$^{2}$~s~nm). The Doppler broadening of the natural $^{229m}$Th linewidth does not violate this assumption. Therefore, the rate of $^{229m}$Th excitation events can be estimated as follows:
\begin{equation}
\frac{d N_{Th229m}}{dt} = n_{Th229} g \frac{\Gamma_\gamma}{2} \lambda_0^2 \frac{dN_\gamma}{dt dE}(E_0)~,
\end{equation}
\noindent where $n_{Th229}$ is the surface density of $^{229}$Th atoms. For the reaction $\gamma + ^{229}Th \to ^{229m}Th$ we have:
$g=1/3$, $\lambda_0\simeq 150$~nm~\cite{Seiferle}, and assuming 1~kBq of $^{229}$Th activity in 1 cm$^2$ film we get $n_{Th229}=3.3\times 10^{14}$ $^{229}$Th atoms/cm$^2$. The resonance decay width depends on the medium refractive index $n$. The theoretical prediction was given by $\Gamma_\gamma \simeq n^3 \times 3\times 10^{-19}$~eV~\cite{Tkalya_gamma}. This can be compared to the existing data on $^{229m}$Th lifetimes $\tau$ in various media:
\begin{equation}
\tau = \frac{1}{\Gamma_\gamma} ~.
\end{equation}
The data shown in Fig.~\ref{fig:lifetime_th229m} demonstrates good agreement with theoretical expectations. Moreover, the ThF$_4$ refractive index at 150~nm is significantly higher than all the measured values~\cite{thf4_refractive_index} and may allow us to verify the theoretical n-dependence with better precision.

Combining all these numbers we obtain the estimate of the excitation rate:
\begin{equation}
\frac{d N_{Th229m}}{dt} \simeq 6.7\times 10^{-14}~nm~ \times n^3  \frac{dN_\gamma}{dt d\lambda}(\lambda_0) ~.
\end{equation}
Using the known photon flux of our VUV-lamp and the ThF$_4$ refractive index $n(150~nm)\simeq 1.826$~\cite{thf4_refractive_index}, we can estimate a $^{229m}$Th production rate of about 1.1~Hz. As the production rate is much faster than the decay rate, the accumulated number of $^{229m}$Th isomers after irradiation time $t_{irr}$ is given by:
\begin{equation}
N_{Th229m} = \frac{dN_{Th229m}}{dt} \tau \Biggl [ 1-e^{-\frac{t_{irr}}{\tau}} \Biggr ] ~.
\end{equation}
\noindent Hence, the number of signal events which can be measured within the time $t_{decay}$ after the irradiation can be written as follows:
\begin{equation}
S_{Th229m} = \frac{dN_{Th229m}}{dt} \tau \Biggl [ 1-e^{-\frac{t_{irr}}{\tau}} \Biggr ] \Biggl \{ 1-e^{-\frac{t_{decay}}{\tau}} \Biggr \} ~.
\end{equation}
The optimal decay time, which maximizes the ratio of the signal to statistical fluctuations of uncorrelated background, is equal to $t_{decay}=\tau$, where $\tau$ is the $^{229m}$Th lifetime. The irradiation time $t_{irr}$ has no optimal value, but we propose to use $t_{irr}=2\tau$ yielding 86\% of saturated activity. Combining these numbers, we can estimate the number of signal events to approximately 217. This value has to be compared to backgrounds due to PMT dark counts, photo- and radio-luminescence. In the comparison, we have to include a PMT quantum efficiency of about 10\%. The PMT acceptance can be estimated from geometrical considerations: the R6835 PMT has an effective diameter of 23~mm, and it is located at 11~mm above the sample, resulting in a maximum angle of 45 degrees. The solid angle fraction, therefore, can be estimated as:
\begin{equation}
\frac{\Omega_{Th229m}}{4\pi} = \sin^2{\frac{\theta_{max}}{2}} = 0.15 ~ .
\end{equation}
However, the angular distribution of the M1 transition is not constant:
\begin{equation}
W(\theta) = 1 + \cos^2{\theta} ~ ,
\end{equation}
enhancing the photon emission in forward and backward directions. The effect of the angular distribution can be estimated by an analytic integral:
\begin{equation}
\frac{\Omega_{Th229m}}{4\pi} = \int_1^{\cos{\theta_{max}}} W(z) dz = \frac{ (1-\cos{\theta_{max}}) + \frac{1}{3}(1-\cos^3{\theta_{max}}) }{2} = 0.27 ~ .
\end{equation}
\noindent Thus, the photon angular distribution of the M1 transition increases the acceptance of our measurement to 72\%. These values can be reduced by the total internal reflection inside ThF$_4$, but the choice of a small enough film thickness should allow recovering a large part of this loss.

Although photo- and radio-luminescences have similar acceptance and efficiency, depending on the number of emitted photons in the VUV range these corrections could be smaller than those required for the signal from $^{229m}$Th decay.

Summarizing, at the VUV PMT we expect $217 \times 0.27 \times 0.1=6$ signal events measured in 360~s, with an average rate of 0.017~Hz and the peak rate of 0.026~Hz. Therefore, the VUV photo-luminescence of the ThF$_4$ background rate should be smaller than this value or the decay time significantly shorter. Instead, for continuous processes (independent from the VUV-light source), like radio-luminescence and PMT dark counting, the variances of both these backgrounds in 360~s should be smaller than 6 events. This means that the event rate of both these backgrounds should be less than 0.1~Hz. The latter condition can be loosened by performing a long series of repetitive measurements. However, in this case, the stability of background rates should be smaller than 0.1~Hz.

\section{Conclusions}\label{sec:conclusions}
In this article, we present the first measurements of photo- and radio-luminescence of ThF$_4$ thin films. The observed photo-luminescence is low in the UV range and not very high in the visible range. Similarly, the observed radio-luminescence is also low in both measured ranges.

These measurements have implications for the potential use of ThF$_4$ thin films enriched with $^{229}$Th to detect $^{229m}$Th decay:

\begin{itemize}
\item the observed UV photo-luminescence of the tested ThF$_4$ films is 8.7 times larger than the expected signal from $^{229m}$Th estimated in section~\ref{sec:projection_expect}.

However, this level of photo-luminescence was observed with a ThF$_4$ sample that had an area of 4.84~cm$^2$ and a thickness of 200~nm. A smaller sample will produce a proportionally smaller amount of photo-luminescence. Specifically, a proposed ThF$_4$ film with an area of 1~cm$^2$ and a thickness of 50~nm is expected to produce 20 times less photo-luminescence. As a result, we anticipate that the UV photo-luminescence of the proposed ThF$_4$ film will be lower than the $^{229m}$Th signal rate. The decay time of the observed photo-luminescence was approximately 400~s, similar to the anticipated $^{229m}$Th decay lifetime in ThF$_4$. Therefore, a comparison with a similar sample containing only natural Th is mandatory. Furthermore, the photo-luminescence is dependent on the quality of the crystal, which can potentially be improved in comparison to that of the utilized commercial samples.
\item The observed visible photo-luminescence is not very large and allows us to use it as a veto for high light-yield luminescence signals. The probability of accidental coincidences with the signal is negligible.
\item the observed UV radio-luminescence of the tested ThF$_4$ films is 0.7~Hz, which is 41 times larger than the signal expected from $^{229m}$Th, as estimated in section~\ref{sec:projection_expect}. This rate is continuous and can be calibrated with the precision given by the statistical deviation of the background. As described in section~\ref{sec:projection_expect}, the continuous background should have an overall rate of less than 0.1~Hz. However, the observed rate is larger and it refers again to the 200~nm thick sample. The probability of producing a UV photon is proportional to the thickness crossed by the $\alpha$-particle. Furthermore, the $\alpha$ particles from the external source used here cross the entire ThF$_4$ film, while intrinsic activity $\alpha$s, on average, will cross only half of it. Therefore, we can rescale the presented measurements to the internal activity of 1~kBq in 50~nm thick ThF$_4$ and obtain 0.2~Hz of internal radio-luminescence background. This is still within an acceptable range for the present proposal, although it may require the measurements to be repeated many times to increase the statistical power.
\item The observed visible radio-luminescence is not very large, which allows us to use it as a veto for high light yield scintillation signals. However, the efficiency of the coincidence veto in the observed radio-luminescence was fairly small.
\end{itemize}

In summary, the measured backgrounds to the future $^{229m}$Th decay experiment in the ThF$_4$ are significant but do not exceed acceptable limits. The measurements indicate that the using of a small and thin ThF$_4$ film allows for reduced backgrounds, while keeping the signal at the same level by increasing the $^{229}$Th enrichment.

\section*{Acknowledgements}
The authors would like to acknowledge the excellent support provided during the experiment by the staff and technical services of JRC (Karlsruhe).
\par
The experimental data used in this research were generated through access to the ActUsLab/PAMEC under the Framework of access to the Joint Research Centre Physical Research Infrastructures of the European Commission (GRaDeTh229 project,  Research Infrastructure Access Agreement N°36345/04.
\bibliographystyle{elsarticle-num}
\bibliography{thf4_lum}

\begin{thebibliography}{10}
\expandafter\ifx\csname url\endcsname\relax
  \def\url#1{\texttt{#1}}\fi
\expandafter\ifx\csname urlprefix\endcsname\relax\def\urlprefix{URL }\fi
\expandafter\ifx\csname href\endcsname\relax
  \def\href#1#2{#2} \def\path#1{#1}\fi

\bibitem{Seiferle}
B.~Seiferle, et~al., {Energy of the 229Th nuclear clock transition}, Nature 573
  (2019) 243.

\bibitem{Tkalya_photoexcitation}
E.~Tkalya, et~al., {{Processes of the nuclear isomer 229mTh(3/2+, 3.5 +- 1.0
  eV) resonant excitation by optical photons}}, Phys.\ Scr. 53 (1996) 296.

\bibitem{peik}
E.~Peik, C.~Tamm, {Nuclear laser spectroscopy of the 3.5 eV transition in
  Th-229}, Europhys. Lett. 61 (2003) 181.

\bibitem{Wense}
L.~{von der Wense}, et~al., {Direct detection of the 229Th nuclear clock
  transition}, Nature 533 (2016) 47.

\bibitem{Seiferle_lifetime}
B.~Seiferle, et~al., {Lifetime Measurement of the 229Th Nuclear Isomer}, Phys.\
  Rev.\ Lett. 118 (2017) 042501.

\bibitem{Tkalya_n3_dep}
E.~Tkalya, {Spontaneous Emission Probability for M1 Transition in a Dielectric
  Medium: 229mTh(3/2+, 3.5 +- 1.0 eV) Decay}, JETP Lett. 71 (2000) 311.

\bibitem{Tkalya_n3_dep_prc}
E.~Tkalya, {Decay of the low-energy nuclear isomer 229mTh(3/2+, 3.5 +- 1.0 eV)
  in solids (dielectrics and metals) A new scheme of experimental research},
  Phys.\ Rev.\ C 61 (2000) 064308.

\bibitem{Yamaguchi_tau_th229m_vacuum}
A.~Yamaguchi, et~al., {Laser spectroscopy of triply charged 229Th isomer for a
  nuclear clock}, Nature 629 (2024) 62.

\bibitem{Tiedau_tau_th229m_caf2}
J.~Tiedau, et~al., {Laser excitation of the Th-229 nucleus}, Phys.\ Rev.\ Lett.
  132 (2024) 182501.

\bibitem{Kraemer_tau_th229m_mgf2}
S.~Kraemer, et~al., {Observation of the radiative decay of the 229Th nuclear
  clock isomer}, Nature 617 (2023) 706.

\bibitem{Elwell_tau_th229m_licaf}
R.~Elwell, et~al., \href{https://arxiv.org/abs/2404.12311}{Laser excitation of
  the 229th nuclear isomeric transition in a solid-state host},
  {arXiv:2404.12311} (2024).
\newline\urlprefix\url{https://arxiv.org/abs/2404.12311}

\bibitem{Tkalya}
E.~Tkalya, {Proposal for a Nuclear Gamma-Ray Laser of Optical Range}, Phys.\
  Rev.\ Lett. 106 (2011) 162501.

\bibitem{band_gap_thf4}
T.~Gouder, et~al., {Measurements of the band gap of ThF$_4$ by electron
  spectroscopy techniques}, Phys. Rev. Research 1 (2019) 033005.

\bibitem{thf4_refractive_index}
W.~Heitmann, E.~Ritter, {Production and properties of vacuum evaporated films
  of thorium fluoride}, Appl. Opt. 7 (1968) 307.

\bibitem{L11798_datasheets}
{Hamamatsu Photonics K. K.},
  \href{https://www.hamamatsu.com/us/en/product/light-and-radiation-sources}{{L11798}
  datasheets} (2022).
\newline\urlprefix\url{https://www.hamamatsu.com/us/en/product/light-and-radiation-sources}

\bibitem{SM05PDA7A_datasheets}
{Thorlabs, Inc.},
  \href{"https://www.thorlabs.com/thorproduct.cfm?partnumber=SM05PD7A"}{{SM05PDA7A}
  datasheets} (2022).
\newline\urlprefix\url{"https://www.thorlabs.com/thorproduct.cfm?partnumber=SM05PD7A"}

\bibitem{R6835_datasheets}
{Hamamatsu Photonics K. K.},
  \href{https://www.hamamatsu.com/us/en/product/optical-sensors/pmt}{{R6835}
  datasheets} (2022).
\newline\urlprefix\url{https://www.hamamatsu.com/us/en/product/optical-sensors/pmt}

\bibitem{ortec113}
\href{https://www.ortec-online.com/products/electronic-instruments/preamplifiers/113}{{ORTEC.
  Available online:}} (2022).
\newline\urlprefix\url{https://www.ortec-online.com/products/electronic-instruments/preamplifiers/113}

\bibitem{DPP-PHA}
\href{https://www.caen.it/products/dpp-pha/}{{CAEN SpA, firmware DPP-PHA.
  Available online:}} (2022).
\newline\urlprefix\url{https://www.caen.it/products/dpp-pha/}

\bibitem{DT5730}
\href{https://www.caen.it/products/dt5730/}{{CAEN SpA, Digitiser DT5730.
  Available online:}} (2022).
\newline\urlprefix\url{https://www.caen.it/products/dt5730/}

\bibitem{CAEN-COMPASS}
\href{https://www.caen.it/products/compass/}{{CAEN SpA, software COMPASS.
  Available online:}} (2022).
\newline\urlprefix\url{https://www.caen.it/products/compass/}

\bibitem{II-VI}
{II-VI Incorporated},
  \href{https://ii-vi.com/wp-content/uploads/2020/03/IR-Optics-Series-web-n.pdf}{Ir
  optics series} (2022).
\newline\urlprefix\url{https://ii-vi.com/wp-content/uploads/2020/03/IR-Optics-Series-web-n.pdf}

\bibitem{uv_in_o2}
K.~Watanabe, et~al., {Absorption coefficients of oxygen in the vacuum
  ultraviolet.}, J. Chem. Phys. 21 (1953) 1026.

\bibitem{geant4}
S.~Agostinelli, et~al., Geant4—a simulation toolkit, Nucl. Instr. and Meth.
  506 (2003) 250.

\bibitem{PDG}
D.~Groom, et~al., {Particle Data Group}, Eur. Phys. Jour. C15 (2000) 1.

\bibitem{Tkalya_gamma}
A.~M. Dykhne, E.~Tkalya, {Matrix element of the anomalously low-energy
  (3.5±0.5 eV) transition in 229Th and the isomer lifetime.}, J. Exp. Theor.
  Phys. Lett. 67 (1998) 251.

\end{thebibliography}


\end{document}